\documentclass[a4paper,11pt]{article}

\usepackage{lmodern}
\usepackage[utf8]{inputenc}
\usepackage[T1]{fontenc}
\usepackage{amsmath}
\usepackage{amssymb}
\usepackage{mathtools}
\usepackage[english]{babel}
\usepackage{url}
\usepackage{booktabs}
\usepackage{longtable}
\usepackage{cellspace}
\usepackage{makecell}
\usepackage{graphicx}
\usepackage{needspace}
\usepackage{xspace}
\usepackage{setspace}
\usepackage{textcase}
\usepackage{multirow}
\usepackage{subcaption}
\usepackage{jheppub}
\usepackage{setspace}
\usepackage{bm}
\usepackage{scrextend}
\usepackage{diagbox}
\usepackage{mathrsfs}
\allowdisplaybreaks[1]

\usepackage[format=plain,labelfont=bf,textfont=it]{caption}

\usepackage[dvipsnames]{xcolor}
\newcommand{\RedTextHTML}{A00000}
\newcommand{\BlueTextHTML}{0B476D}
\definecolor{RedText}{HTML}{\RedTextHTML}
\definecolor{BlueText}{HTML}{\BlueTextHTML}
\definecolor{FittingYellow}{HTML}{ADAB11}
\definecolor{BGColorL}{HTML}{F3F3F3}
\definecolor{BGColorH}{HTML}{D9D9D9}

\usepackage{enumitem}
\setlist[description]{%
	itemsep=0pt,               
	font={\normalfont\scshape}, 
}

\usepackage{tikz}
\usetikzlibrary{intersections}
\usetikzlibrary{decorations.markings}
\usetikzlibrary{arrows.meta}
\usetikzlibrary{calc}
\usetikzlibrary{graphs}

\makeatletter
\global\let\tikz@ensure@dollar@catcode=\relax
\makeatother

\newcommand{\mycomment}[1]{}

\let\originalleft\left
\let\originalright\right
\renewcommand{\left}{\mathopen{}\mathclose\bgroup\originalleft}
\renewcommand{\right}{\aftergroup\egroup\originalright}

\makeatletter
\providecommand*{\shuffle}{%
	\mathbin{\mathpalette\myshuffle@{}}%
}
\newcommand*{\myshuffle@}[2]{%
	\sbox0{$#1\vcenter{}$}%
	\kern .15\ht0 
	\rlap{\vrule height .25\ht0 depth 0pt width 2.5\ht0}%
	\raise.1\ht0\hbox to 2.5\ht0{%
		\vrule height 1.75\ht0 depth -.1\ht0 width .17\ht0 %
		\hfill
		\vrule height 1.75\ht0 depth -.1\ht0 width .17\ht0 %
		\hfill
		\vrule height 1.75\ht0 depth -.1\ht0 width .17\ht0 %
	}%
	\kern .15\ht0 
}
\makeatother

\DeclareFontFamily{U}{mathx}{\hyphenchar\font45}
\DeclareFontShape{U}{mathx}{m}{n}{
	<5> <6> <7> <8> <9> <10>
	<10.95> <12> <14.4> <17.28> <20.74> <24.88>
	mathx10
}{}
\DeclareSymbolFont{mathx}{U}{mathx}{m}{n}
\DeclareFontSubstitution{U}{mathx}{m}{n}
\DeclareMathAccent{\widecheck}{0}{mathx}{"71}

\renewcommand{\Re}{\operatorname{Re}}

\newcommand{\ym}{\text{YM}}

\newcommand{\RR}{\mathbb{R}}
\newcommand{\NN}{\mathbb{N}}
\newcommand{\CC}{\mathbb{C}}
\newcommand{\QQ}{\mathbb{Q}}
\newcommand{\ZZ}{\mathbb{Z}}

\newcommand{\SL}{\mathrm{SL}}

\newcommand{\SO}{\mathrm{SO}}

\newcommand{\HH}{\mathbb{H}}

\newcommand{\EE}{\mathrm{E}}
\newcommand{\GG}{\mathrm{G}}

\DeclareMathOperator{\KN}{KN}

\newcommand{\ap}{\alpha'}

\newcommand{\cformp}[1]{\operatorname{\mathcal{C}^+\hspace{-3pt}}\big[
	\protect\begin{smallmatrix}#1\protect\end{smallmatrix}\parbox{4pt}{$\big]$}}

\newcommand{\ompm}[3]{\omega_{\pm}\! \left[\begin{smallmatrix}#1\\#2\end{smallmatrix};#3\right]}
\newcommand{\omplus}[3]{\omega_{+}\! \left[\begin{smallmatrix}#1\\#2\end{smallmatrix};#3\right]}
\newcommand{\omminus}[3]{\omega_{-}\! \left[\begin{smallmatrix}#1\\#2\end{smallmatrix};#3\right]}

\newcommand{\beqv}[2]{\beta^{\rm eqv}\! \left[\begin{smallmatrix}#1\\#2\end{smallmatrix}\right]}
\newcommand{\eqv}{\text{eqv}}

\newcommand{\betaeqv}[1]{
	\beta^\mathrm{eqv}\! \big[\begin{smallmatrix}#1\end{smallmatrix}\big]}

\newcommand{\betaeqvtau}[1]{
	\beta^\mathrm{eqv}\! \big[\begin{smallmatrix}#1\end{smallmatrix}; \tau\big]}
\newcommand{\betaeqvx}[2]{
	\beta^\mathrm{eqv}\! \big[\begin{smallmatrix}#1\end{smallmatrix}; #2\big]}

\newcommand{\betaplus}[1]{
	\beta_\mathrm{+}\! \big[\begin{smallmatrix}#1\end{smallmatrix}\big]}
\newcommand{\betaplustau}[1]{
	\beta_\mathrm{+}\! \big[\begin{smallmatrix}#1\end{smallmatrix}; \tau\big]}

\newcommand{\betaminus}[1]{
	\beta_\mathrm{-}\! \big[\begin{smallmatrix}#1\end{smallmatrix}\big]}
\newcommand{\betaminustau}[1]{
	\beta_\mathrm{-}\! \big[\begin{smallmatrix}#1\end{smallmatrix}; \tau\big]}

\newcommand{\cforml}[1]{\operatorname{\mathcal{C}\hspace{-3pt}}\Big[
	\protect\begin{smallmatrix}#1\protect\end{smallmatrix}\parbox{4pt}{$\Big]$}}

\newdimen\aboverulesepbuffer
\newdimen\belowrulesepbuffer

\title{\boldmath Five-point Type IIB String Amplitudes at One Loop}

\author[a]{E. Claasen}
\author[b]{M. Doroudiani}
\affiliation[a]{Max-Planck-Institut f\"ur Gravitationsphysik (Albert-Einstein-Institut), Am M\"uhlenberg 1, DE-14476 Potsdam, Germany}
\affiliation[b]{School of Physics \& Astronomy, University of Southampton, SO17 1BJ, UK}

\emailAdd{emiel.claasen@aei.mpg.de}
\emailAdd{m.doroudiani@soton.ac.uk}

\abstract{Massless type IIB superstring amplitudes are organized according to the number of external states and their ${\rm U}(1)$ charge under the R-symmetry of type IIB supergravity. In this work, we analyze the low-energy expansion of one-loop five-point amplitudes in all charge sectors, focusing on the representative processes involving five gravitons and four gravitons with one dilaton. We compute the one-loop contributions to the moduli-dependent couplings in the type IIB effective action up to the $D^{12}R^5$ and $D^{14}\phi R^4$ interactions. The results are consistent with $S$-duality constraints in every charge sector and exhibit rich arithmetic structure, including single-valued multiple zeta values, affine linear combinations of logarithmic derivatives of the Riemann zeta function at odd integers, and a new constant of currently unknown nature.}

\makeatletter
\gdef\@fpheader{}
\makeatother

\begin{document}
\maketitle
\flushbottom

\section{Introduction}
\label{sec:intro}
Progress in field theory amplitude computations can be measured through the two axes of number of loops and number of external states. In string theory a third axis emerges: whenever the external momenta in a string scattering process are small compared to the inverse square root of the string scale $\ap$, one may set up a perturbative expansion in the momenta. This is known as the low-energy expansion of string amplitudes, or equivalently, as the $\ap$- or field-theory expansion. Low-energy expansions of superstring amplitudes have exhibited fruitful synergies with methods from number theory and algebraic geometry, a prominent example being the appearance of (single-valued) multiple zeta values as expansion coefficients in the calculation of type I/II , heterotic and bosonic string amplitudes at tree level \cite{Schlotterer:2012ny,Stieberger:2013wea,Stieberger:2014hba,Schlotterer:2018zce,Vanhove:2018elu,Brown:2019wna,Huang:2016tag,Azevedo:2018dgo}.

In addition to the dependence on the number of external states, the specific kind of external states also plays a role. Focusing on the scattering of massless states in type IIB string theory, the external states are captured in a supermultiplet containing particles with different amounts of charge under the ${\rm U}(1)$ R-symmetry of the underlying type IIB supergravity. R-symmetry is not a symmetry of string theory, which means that processes involving external states whose combination of R-charges is non-zero are allowed. These are known as ${\rm U}(1)$-violating amplitudes. An important property is that amplitudes with the same total amount of R-charge are related by supersymmetry. This reduces the study of all possible amplitudes to one representative of each R-charge sector. Furthermore, for a process involving $N$ external states, the maximal amount of R-charge is bounded by $\lvert q\rvert_{\rm max}=2N-8$ \cite{Green}. This means that at four points, there is only one representative process with zero R-charge, whereas at five points we open up the possibility of having non-zero R-charge. Amplitudes that saturate the R-charge bound are called maximally R-symmetry violating (MRV) amplitudes and were shown to exhibit remarkable simplicity \cite{Boels:2012zr,Boels:2013jua,Green:2013bza} akin to the simplicity of maximally helicity violating amplitudes in Yang-Mills theory \cite{Parke:1986gb,Berends:1987me}.

In this work, we study perturbative contributions to the low-energy effective action in type IIB superstring theory that arise from five-point scattering processes in all possible charge sectors. Perturbative closed-string amplitudes are expressed as series expansions in small values of the string coupling $g_s$. The kinematically-dependent coefficients in such expansions are integrals over the moduli spaces of punctured Riemann surfaces associated to the closed-string worldsheet of the specific process. These integrals can be split into integrals over the complex structure moduli and over the configuration space of the punctures, where we consider the latter integral to be the \textit{integrand} of the former. Calculating an amplitude for a particular scattering process amounts to evaluating these types of integrals. 

In recent years, substantial progress has been made in elucidating the structure of the low-energy expansion of these integrals at genus one. In particular, the low-energy expansion of the genus-one integrand defines a class of non-holomorphic $\SL(2,\ZZ)$ modular forms. These were originally studied using the formalism of modular graph forms (MGFs), which provides graphical representations of integrand contributions and Feynman-like rules for constructing $\SL(2,\ZZ)$-covariant lattice sums \cite{Green:1999pv,Green:2008uj,DHoker:2015gmr,DHoker:2015wxz,DHoker:2015sve,DHoker:2016mwo,DHoker:2016quv,Basu:2016kli,DHoker:2017zhq,Kleinschmidt:2017ege,Gerken:2018zcy,Gerken:2020aju}. However, intricate relations among MGFs, the absence of a canonical basis, and their poor suitability for performing the remaining moduli integrals have made higher orders in the $\alpha'$ expansion difficult to access \cite{DHoker:2019blr,DHoker:2019mib,Doroudiani:2023bfw,Claasen:2024ssh,Basu:2016mmk}.

The first two obstacles have been resolved with the advent of the equivariant iterated Eisenstein integral (EIEI) formalism. EIEIs are defined through an infinite family of non-holomorphic modular forms constructed from iterated integrals of holomorphic Eisenstein series \cite{Brown:mmv,Brown:2017qwo,Brown:2017qwo2,Broedel:2018izr,Gerken:2019cxz,Gerken:2020yii,Gerken:2018jrq,Dorigoni:2022npe,Dorigoni:2024oft,Claasen:2025vcd}, where an initial connection to MGFs was identified by matching Laplace equations. Since then, the translation of MGFs into EIEIs has been made completely algorithmic \cite{Claasen:2025vcd}. Since EIEIs are algebraically independent \cite{Matthes_2017} and admit a canonical basis\footnote{To be more precise, the space of MGFs is a proper subspace of EIEIs; consequently, the canonical basis for MGFs can only be obtained by projecting onto this specific subspace. The linear combinations of EIEIs that characterize the MGF basis are governed by the Pollack relations \cite{Brown:2017qwo2,Dorigoni:2024oft}. We discuss this in further detail in section \ref{sec:oneloop}.}, they provide a more natural function space than MGFs. Still, the main bottleneck of evaluating the amplitude is the remaining integration over the moduli space of the torus. For the function space defined by EIEIs, this is under control up to iteration depth 3 \cite{Doroudiani:2023bfw}, which limits the order in the $\ap$ expansions. Also, see \cite{DHoker:2019mib,DHoker:2021ous} for the integration of the lattice-sum representation of MGFs for two- and three-loop graphs, which are subclasses of EIEIs at depth 2 and 3. As of the time this paper was written, no MGF containing an EIEI of depth four or higher has been integrated.

We focus on calculating five-point processes in type IIB string theory that contribute to the low-energy effective action up to $D^{12}R^5$ and $D^{14}\phi R^4$ for the ${\rm U}(1)$-conserving and (maximally) violating sectors respectively (for more developments of the five-point effective action, see \cite{Richards:2008jg,Liu:2022bfg,Liu:2025uqu}). The kinematic matrices that arise in the conserving sector partially mimic that of the five-graviton tree-level amplitude and also turn out to be closely related to the one-loop four-graviton amplitude, which extends beyond the results in \cite{Green:2013bza,Basu:2016mmk}. Furthermore, the same matrices arise in the violating sector with some minor modifications that can be explained from $S$-duality. Beyond these structures, we find genuine five-point operators that do not factorize onto four-point processes. The coefficients of all these kinematic matrices involve number-theoretically interesting combinations of multiple zeta values, logarithmic derivatives of zeta values, the Euler-Mascheroni constant, and a new number whose nature we were unable to identify and we only determined numerically. A striking pattern among the numbers in these coefficients is that the logarithmic derivatives of zeta values and the Euler-Mascheroni constant arise in an affine linear combinations (meaning the coefficients add up to one) in all examples, which is a strong indication of an underlying mathematical framework that we were not able to identify.

\subsection{Overview}
This paper is organized as follows. In section \ref{section:review}, we review the basic building blocks of one-loop closed-string scattering amplitudes. In particular, we review notions of $S$-duality and R-symmetry in type IIB in \ref{ssec: sRtypeIIB}. In sections \ref{sec:tree} and \ref{sec:oneloop}, we discuss the building blocks of closed string amplitudes at both tree level and one loop respectively. In section \ref{sec:gen1}, we specialize to the scattering of five external states at one loop. As a warm-up, we first go through the calculation of the four-graviton one-loop amplitude in section \ref{ssec:4ptrev}, which plays an important role in the later sections. The general formulas that are valid in all R-charge sectors for five-point processes are discussed in \ref{ssec:gen5pt}. This is used in sections \ref{ssec: conserv} and \ref{ssec:viol} to calculate one-loop contributions to the effective action up to $D^{12}R^5$ and $D^{14}\phi R^4$ respectively. These results are connected through $S$-duality as shown in \ref{ssec: sdual}. Finally in \ref{ssec: comments} we comment on general patterns observed in the results of both four- and five-point superstring scattering processes in type IIB superstring theory. The arXiv submission of this work comes with a Mathematica notebook containing all the relevant five-point kinematic matrices, the moduli space integrands and their integrals. Furthermore, two text files are supplied that realize a change of basis between modular invariant integrands in order to adhere to the basis used in \cite{Doroudiani:2023bfw}.

\section{Review of genus-zero and genus-one string amplitudes}
\label{section:review}
In this section, we collect the basic ingredients needed for the study of five-point genus-one string amplitudes in type IIB string theory. Central to our discussion is $S$-duality and its interplay with R-symmetry, reviewed in section~\ref{ssec: sRtypeIIB}. Section~\ref{sec:tree} provides a review of massless genus-zero amplitudes, which serves as a precursor to the genus-one story presented in section~\ref{sec:oneloop}. Our treatment of genus-one amplitudes employs the formalism of modular graph forms and their representation in terms of iterated Eisenstein integrals. The conversion algorithm between these two languages is reviewed in section~\ref{ssec:conv}.

\subsection{$S$-duality and R-symmetry in type IIB}
\label{ssec: sRtypeIIB}
The material presented in this section is mainly based on \cite{Fleig:2015vky,Green:2013bza}. The moduli space of classical type IIB supergravity in ten-dimensional flat spacetime is the symmetric space ${\rm SL}(2,\mathbb R)/{\rm SO(2,\mathbb R)}$. In type IIB superstring theory, the classical moduli space gets quantized to 
\begin{align}
	\mathcal{M}_{\text{quantum}}=\SL(2,\ZZ)\backslash\SL(2,\RR)/\SO(2,\RR)\,,
	\label{eq: qmodspace}
\end{align}
and is characterized by the axio-dilaton, which can be written as a complex scalar $\Omega=C^{(0)}+ie^{-\varphi_s}=:\Omega_1+i\Omega_2$, with $C^{(0)}$ the Ramond-Ramond zero-form and $\varphi_s$ the dilaton. The discrete group $\SL(2,\ZZ)$ identifies string theories at different points in the quantum moduli space, which is known as $S$-duality \cite{Hull:1994ys}. It acts on the axio-dilaton via fractional linear transformations
\begin{align}
	\Omega\mapsto g \Omega:=\frac{\alpha\Omega+\beta}{\gamma\Omega+\delta}\,,\qquad g= \begin{pmatrix}
		\alpha&\beta\\\gamma&\delta
	\end{pmatrix}\in {\rm SL}(2,\mathbb Z)\,.
	\label{eq: sl2act}
\end{align}
Any physical observable in string theory, such as a scattering amplitude, must be invariant under this duality group, in a sense to be made precise below \eqref{eq: fieldtransf}. If we additionally require perturbative behavior in the weak coupling limit $g_{s}=e^{\varphi_s}\rightarrow 0$, this leads us to the realm of modular forms.

Holomorphic modular forms of weight $k$ are holomorphic functions $f:\HH\rightarrow \CC$ where $\HH=\{\Omega\in\CC\mid {\rm Im}\,\Omega>0\}$ that transform under the action of $\SL(2,\mathbb Z)$ as 
\begin{align}
	f(g \Omega)=(\gamma\Omega+\delta)^kf(\Omega)\,.
	\label{eq:holmod}
\end{align}
and they have suitable growth near the cusp $\Omega\rightarrow i\infty$ \cite{serre2012course}. Their real-analytic variants are known as non-holomorphic modular forms. These are real-analytic functions $f:\HH\rightarrow\CC$ that transform as
\begin{align}
	f(g\Omega)=(\gamma \Omega+\delta)^a(\gamma\overline{\Omega}+\delta)^bf(\Omega)\,,
	\label{eq:nonholmod}
\end{align}
with moderate (polynomial) growth at the cusp. We call $(a,b)$ the holomorphic and anti-holomorphic weights respectively, and $a+b$ the total modular weight. 

Type IIB superstring perturbation theory in ten-dimensional flat space is an expansion around a limit in moduli space in which the string coupling $g_{s}=e^{\varphi_s}$ goes to zero. However, the amplitude is not expected to be a convergent series in the string coupling $g_{s}$, meaning that there are non-perturbative effects arising from instanton configurations \cite{Green:1997tv,Pioline:2009ia,Shenker:1990uf}. Instead of setting up a perturbation theory in the string coupling, one can write down a low-energy effective action that captures the stringy corrections that are analytic in $\ap$. Contributions to the effective action arise from scattering processes represented by local operators $\mathcal{O}_q$ constructed from (derivatives of) type IIB fields with total ${\rm U}(1)$-charge $q$. The fields $\Phi_q$ in the massless type IIB supergravity multiplet transform under $S$-duality as non-holomorphic modular forms with weights corresponding to the ${\rm U}(1)$-charge of the R-symmetry group acting on the supermultiplet 
\begin{align}
	\Phi_q(g \Omega)=\bigg(\frac{\gamma\Omega+\delta}{\gamma\overline{\Omega}+\delta}\bigg)^{q/2}\Phi_q(\Omega)\,.
	\label{eq: fieldtransf}
\end{align}
The ${\rm U}(1)$-charges $q$ of the fields in the supermultiplet can be found in table \ref{tab:supmultq}.
\begin{table}[]
	\centering
	\begin{tabular}{|c|c|c|c|}
		\hline \textbf{boson} & $q$ & \textbf{fermion} & $q$ \\
		\hline (anti-)holomorphic axio-dilaton & $\pm 2$ & dilatino & $\pm 3/2$\\
		(anti-)holomorphic three-form & $\pm 1$ & gravitino & $\pm 1/2$ \\
		graviton and five-form &0&&\\\hline
	\end{tabular}
	\caption{Type IIB supermultiplet states with their ${\rm U}(1)$-symmetry charges $q$.}
	\label{tab:supmultq}
\end{table}
As the fields and thus the operator $\mathcal{O}_q$ transform under $S$-duality, this must be compensated for by additional functions of the moduli $f^{(-q,q)}(\Omega)$ such that the total contribution is $\SL(2,\ZZ)$-invariant. The transformation property of the fields (\ref{eq: fieldtransf}) forces the $f^{(-q,q)}(\Omega)$ to transform as a non-holomorphic modular form of weights $(-q,q)$, so that individual terms in the effective action take the schematic form
\begin{align}
	S\supset\int {\rm d}^{10}x\, (\ap)^p f^{(-q,q)}(\Omega)\mathcal{O}_q\,,
	\label{eq: effact}
\end{align}
where the power $p$ of $\ap$ can be fixed by dimensional analysis.

The ${\rm U}(1)$ R-symmetry is a symmetry of the type IIB supergravity effective action, but not of the full type IIB string theory. Accordingly, string theory admits amplitudes with a nonzero total 
${\rm U}(1)$-charge, which may be viewed as ${\rm U}(1)$-violating from the supergravity perspective and hence should vanish in the $\ap\rightarrow 0$ limit. One might be tempted to choose a family of external states from the massless type IIB multiplet in table \ref{tab:supmultq} and study its amplitude and contribution to the effective action. However, many of such amplitudes either vanish or are related by supersymmetry. In fact, amplitudes whose external states have the same total ${\rm U}(1)$ charge are related by supersymmetry. Furthermore, amplitudes of multiplicity $N$ whose charge exceed the bound 
\begin{align}
	\lvert q\rvert_{\rm max}=2N-8\,,
	\label{eq: qmax}
\end{align}
are zero \cite{Green}. 

As an example, according to (\ref{eq: qmax}), the only nonzero four-point amplitudes must have $q=0$. Taking the scattering of four gravitons as a representative process, we find the following effective action\footnote{The subscript notation of the coefficient functions is based on \cite{Green:2013bza} and can differ from other sources. The reason for this notation will become clear in section \ref{sec:gen1}.} 
\begin{align}
	S=S_{\text{class.}}+\int d^{10}x\sqrt{-G}&\Big((\ap)^3\mathcal{E}_{3}(\Omega)R^4+(\ap)^5\mathcal{E}_{5}(\Omega)D^4R^4+(\ap)^6\mathcal{E}_{3,3}(\Omega)D^6R^4\nonumber\\&+(\ap)^7\mathcal{E}_{7}D^8R^4 +(\ap)^8\mathcal{E}_{\{3,5\}}D^{10}R^4+\mathcal{O}\big ((\ap)^9\big)\Big)\,,
	\label{eq: effaction}
\end{align}
where $S_{\text{class.}}$ is the classical, zeroth-order in $\ap$, effective action described by type IIB supergravity. The $D$ are covariant derivatives and $R^4$ is a contraction of two $t_8$ or two $\epsilon_{10}$ tensors and four Riemann curvature tensors \cite{Richards:2008jg}. As the total ${\rm U}(1)$-charge of the operators is zero, the coefficients $\mathcal{E}_{\bullet}(\Omega)$ are non-holomorphic modular forms of weights $(0,0)$ and contain both perturbative and non-perturbative contributions in $g_s$. 
Generally, the coefficient functions $\mathcal{E}_{\bullet}(\Omega)$ are not known, but there are some results on the first few corrections to the classical supergravity action in the above example. The first and second corrections arise from $1/2-$BPS and $1/4-$BPS interactions respectively. Supersymmetry forces these coefficient functions to be solutions of Laplace equations \cite{Green:1997tv,Green:1997tn,Green:1998by,Pioline:2001jn}
\begin{align}
	\Delta_\Omega \mathcal{E}_3(\Omega) =\frac{3}{4}\mathcal{E}_3(\Omega)\,,\qquad  \Delta_\Omega \mathcal{E}_5(\Omega) =\frac{15}{4}\mathcal{E}_5(\Omega)\,,
	\label{eq: diffeqns}
\end{align}
with $\Delta_\Omega:=4\Omega_2^2\partial_\Omega\partial_{\overline{\Omega}}$. The unique ${\rm SL}(2,\mathbb Z)$-invariant solutions that are power behaved when $ \Omega_2\rightarrow \infty $ are the real-analytic Eisenstein series
\begin{align}
	\mathcal{E}_3(\Omega)=\pi^{3/2}\EE_{3/2}(\Omega)\,,\qquad\mathcal{E}_{5}(\Omega)=\frac{\pi^{5/2}}{2}\EE_{5/2}(\Omega)\,.
	\label{eq: sols}
\end{align}
They have a Fourier expansion
\begin{align}
	\EE_s(\Omega)= \sum_{n=0}^\infty a_n(\Omega_2)e^{2\pi in\Omega_1}\,.
	\label{eq: realaneis}
\end{align}
The zero modes $a_0(\Omega_2)$ of the Fourier expansions of (\ref{eq: sols}) contain information about the perturbative contributions in the string coupling $g_s$. In particular, they are given by
\begin{align}
	\EE_s(\Omega)=2\pi^{-s}\zeta_{2s}\Omega^s_2+\frac{2\pi^{1/2-s}\Gamma(s-1/2)}{\Gamma(s)}\zeta_{2s-1}\Omega_2^{1-s}+\text{non-zero modes}\,,
	\label{eq: nonholeisom}
\end{align}
and can be identified with (\ref{eq: effaction}) by transforming (\ref{eq: nonholeisom}) from Einstein frame to string frame with contributions coming from tree-level and one-loop amplitudes for $s=3/2$, and from tree-level and two-loop amplitudes for $s=5/2$. These have been explicitly checked by perturbative amplitude computations in \cite{DHoker:2005jhf,Gomez:2010ad}. The non-zero modes contain the effects of $D$-instantons. For a detailed treatment of $D$-instanton effects, we refer the reader to, e.g., \cite{Sen:2020cef, Sen:2021tpp, Agmon:2022vdj}. The modular function $\mathcal{E}_{3,3}(\Omega)$ associated to the $D^6R^4$ correction is also known and is given by a generalized Eisenstein series \cite{Green:2014yxa}. For discussions on coefficient functions of higher order corrections, we refer to \cite{Green:2005ba,Ahlen:2018wng,Dorigoni:2019yoq,Green:2013bza,Green:2008uj}.

Advancing from four- to five-point amplitudes, we open up the possibility of ${\rm U}(1)$-charge violation by a maximum of $\lvert q\rvert_{\rm max}=2$ according to (\ref{eq: qmax}). In \cite{Green:1998by,Green:1997me} it was shown that the coefficient functions of the leading interactions in case of maximal violation are related by nonlinear supersymmetry to $(\ap)^3 \mathcal{E}_3(\Omega)R^4$. For example, the five-point amplitude with $q=-2$ built from one anti-holomorphic dilaton $\phi$ and four gravitons has leading coefficient function 
\begin{align}
	\nabla^{(0)}\mathcal{E}_3(\Omega)=\sum_{(m,n)\in\ZZ^2}^\prime \frac{\Omega_2^{3/2}}{\lvert m\Omega+n\rvert^3}\bigg(\frac{m\overline{\Omega}+n}{m\Omega+n}\bigg)\,,
	\label{eq:derivE3}
\end{align}
where the prime over the summation sign means that the summation excludes $(m,n)=(0,0)$ and $\nabla^{(a)}=2i\Omega_2\partial_\Omega+a$ is a modular covariant derivative acting on the space of non-holomorphic modular forms of weights $(a,b)$ and maps it to forms of weights $(a+1,b-1)$. The conjugate operator $\overline{\nabla}^{(b)}=-2i\Omega_2\partial_{\overline\Omega}+b$ maps from forms of weights $(a,b)$ to $(a-1,b+1)$. They are known as Maass operators and can be interpreted as $\SL(2,\mathbb R)$ raising and lowering operators. The zero mode of (\ref{eq:derivE3}) is obtained by acting with $\nabla^{(0)}$ on the zero mode of $\mathcal{E}_3(\Omega)$, which allows us to relate the perturbative behavior of the maximally R-symmetry violating five-point amplitude to the four-point amplitude. In section \ref{ssec: sdual}, we will see that this pattern continues for higher derivative corrections.

\subsection{Massless genus-zero superstring amplitudes}
\label{sec:tree}
In this work, we focus on the computation of perturbative one-loop contributions to the coefficient functions in the effective action. For this reason, in this section we switch to the framework of perturbative superstring amplitudes and only do a low-energy expansion afterwards. Before proceeding to genus-one superstring amplitudes, it is instructive to look at their genus-zero counterparts. Furthermore, since open-string tree-level amplitudes act as building blocks of closed-string genus-zero amplitudes through the Kawai-Lewellen-Tye (KLT) procedure \cite{Kawai:1985xq}, it is natural to introduce them first. In this part of the review, we follow \cite{Green:2013bza}.

Superstring amplitudes are functions of the external momenta and polarizations.  Tree-level amplitudes for any number $N$ of massless open strings have been computed using pure-spinor cohomology methods in \cite{Mafra:2011nv,Mafra:2011nw}. Writing the result in the $(N-3)!$-dimensional basis of color-stripped amplitudes \cite{BjerrumBohr:2009rd, Stieberger:2009hq}, this gives
\begin{align}
	\mathcal{A}_{\rm{tree}}\big(1,\sigma(2,3,\dots,N-2),N-1,N\big)=\sum_{\pi\in S_{N-3}}A^\pi_{\rm YM}F_{\rm tree}^{\sigma,\pi}(s_{ij})\,,
	\label{eq: opstring}
\end{align}
where $\pi$ and $\sigma$ are the elements of the group $S_{N-3}$, the permutation of $N-3$ points. This equation is valid in any spacetime dimension, for any compactification, and any amount of supersymmetry. The polarization dependence is completely captured in the massless super Yang-Mills field theory amplitudes $A_{\rm YM}^\pi$ for all processes simultaneously \cite{Mafra:2011nv}. A scattering process with a specific choice of external gluons and gluinos can be selected by projecting the superspace expression to a specific component. The open string amplitude can be viewed as an extension of the massless supersymmetric Yang-Mills amplitudes $A^\pi_{\rm YM}$ for any multiplicity $N$, which appear in their $(N-3)!$ dimensional basis \cite{loopBCJ}
\begin{align}
	A_{\rm YM}^\pi:=A_{\rm YM}(1,\pi(2,3,\dots,N-2),N-1,N)\,.
	\label{eq: aymbasis}
\end{align}
The universal stringy part of the amplitude is encoded in the $F_{\rm tree}^{\sigma,\pi}(s_{ij})$ arising from worldsheet integrals over the disk boundary and are so-called Euler or Selberg integrals given by \cite{Mafra:2011nv}
\begin{align}
	F_{\rm tree}^{\sigma,\pi}(s_{ij})=4^{N-3}\int_{z_{\sigma(i)}<z_{\sigma(i+1)}}\!\!\!\!\!\!\!\!\!\!\!\!\!\!\!\!\!\!\text{d}z_2\ldots\int \text{d}z_{N-2}\prod_{i<j}^{N-1}\lvert z_{ij}\rvert^{4s_{ij}}\pi\left\{\prod_{k=2}^{\lfloor N/2\rfloor}\sum_{m=1}^{k-1}\frac{s_{mk}}{z_{mk}}\prod_{k=\lfloor N/2\rfloor+1}^{N-2}\sum_{n=k+1}^{N-1}\frac{s_{kn}}{z_{kn}}\right\}\,,
	\label{eq: eulerselberg}
\end{align}
where the conformal Killing group $\rm SL(2,\mathbb{R})$ of the disk has been employed to fix three of the worldsheet positions to $(z_1,z_{N-1},z_N)=(0,1,\infty)$. In addition, we defined\footnote{To adhere to the conventions of \cite{Green:2013bza}, the Mandelstam variables differ by a minus sign compared to the recent one-loop four-graviton calculations \cite{Claasen:2024ssh}.}
\begin{align}
	z_{ij}:=z_i-z_j\,,\qquad s_{ij}:=\frac{\alpha^\prime}{4}(k_i+k_j)^2\,,
	\label{eq: mandels}
\end{align}
with $k_i$ the external momenta of the strings that satisfy the on-shell condition $k_i^2=0$ and momentum conservation. 

The disk integrals in (\ref{eq: eulerselberg}) encode the full $\alpha^\prime$-dependence of the string amplitude through the dependence on the $s_{ij}$ invariants. Expanding the integrals in $s_{ij}$ for small momenta, we get an expansion of (\ref{eq: eulerselberg}) organized by orders in $\ap$, known as the low-energy expansion or the $\ap$-expansion. The coefficients in this expansion involve multiple zeta values (MZVs) \cite{Schlotterer:2012ny,Mafra:2011nw,Stieberger:2009rr, Brown:2009qja,Terasoma} defined as 
\begin{align}
	\zeta_{n_1,\ldots,n_r}:=\sum_{0<k_1<\ldots<k_r}\prod_{l=1}^r k_l^{-n_l}\,,
	\label{eq: mzvs}
\end{align}
for $n_l\in\NN$ and $n_r\geq 2$. Here, $r$ is the depth of the MZV and $w=\sum_{i=1}^rn_i$ its weight. The weight coincides with the expansion order shifted by three units, which is a property known as uniform transcendentality \cite{Schlotterer:2012ny}. It is convenient to introduce some notation for the coefficients of even and odd zeta values in the low-energy expansion of the matrix (\ref{eq: eulerselberg}) as\footnote{These coefficients depend on the choice of a $\mathbb{Q}$-basis of MZVs at each weight. We use the conjectural basis defined in \cite{Blumlein:2009cf}.}
\begin{align}
	P_{2k}:=F_{\text{tree}}(\tfrac{s_{ij}}{4})\rvert_{\zeta_{2}^k}\,,\qquad M_{2k+1}:=F_{\text{tree}}(\tfrac{s_{ij}}{4})\Big\rvert_{\zeta_{2k+1}}\,, 
	\label{eq: mmatrix}
\end{align}
whose entries by uniform transcendentality are degree $2k$ and $2k+1$ polynomials in the Mandelstam variables respectively. Explicit examples up to certain order in $\ap$ and multiplicity $N$ can be found in \cite{MZVWebsite}. The explicit low-energy expansion of the amplitude (\ref{eq: opstring}) in terms of MZVs becomes rather unwieldy at high orders in $\ap$, see section \ref{ssec:exmpls} and \cite{Schlotterer:2012ny}. Alternatively, one can use motivic multiple zeta values $\zeta^{\mathfrak{m}}_{n_1,\ldots, n_r}$ \cite{Brown:2011ik} that conjecturally satisfy the same relations as the multiple zeta values in (\ref{eq: mzvs}) to write a motivic lift of the amplitude. Using the Hopf-algebra isomorphism $\phi$ to map motivic MZVs to the $f$-alphabet \cite{Brown:2011ik}, we get a very clean presentation of the expansion of the integrals (\ref{eq: eulerselberg}), which is valid for any multiplicity $N$ \cite{Schlotterer:2012ny,Brown:2011ik,Broedel:2013aza,Drummond:2013vz}
\begin{align}
	F_{\text{tree}}^\mathfrak{m}(s_{ij})=\phi^{-1}\bigg\{\bigg(\sum _{k=0}^\infty f_2^{k}P_{2k}\bigg)\bigg(\sum_{p=0}^\infty\sum_{\substack{i_1,\ldots,i_p\in2\mathbb{N}+1}}f_{i_1}f_{i_2}\ldots f_{i_p} M_{i_p}\ldots M_{i_2} M_{i_1}\bigg)\bigg\} \,,
	\label{eq: ftree}
\end{align}
with $P_0$ being the identity matrix. The motivic nature of the expression is indicated with the superscript $\mathfrak{m}$, not to be confused with the matrix index-notation from before.\\
\indent Using the color-stripped open-string results, we can construct the closed string amplitude through the KLT relations \cite{Kawai:1985xq}. Closed-string correlation functions at tree level factorize into left- and right-moving correlators, which means that the open-string expression (\ref{eq: opstring}) extends to the closed-string tree-level result as 
\begin{align}
	\mathcal{M}_{\text{tree}}=\sum_{\sigma,\pi\in S_{N-3}}A_{\rm YM}^\sigma \mathcal{S}_{\text{tree}}^{\sigma,\pi}(s_{ij})\tilde{A}_{\rm YM}^\pi\,,
	\label{eq: clstring}
\end{align}
where $\tilde{A}_{\rm YM}$ denotes another copy of the Yang-Mills tree amplitude, with the tilde referring to the right-moving degrees of freedom on the string worldsheet. In the same way as before, the super Yang-Mills field theory amplitudes capture all polarization information simultaneously. The $\mathcal{S}_{\text{tree}}^{\sigma,\pi}(s_{ij})$ forms a $(N-3)!\times (N-3)!$ matrix of worldsheet integrals over the sphere that encode the stringy part of the amplitude and is given by \cite{Stieberger:2009hq,Green:2013bza}
\begin{align}
	\mathcal{S}_{\text{tree}}^{\sigma,\pi}(s_{ij}):=\int d^2z_2\ldots\int d^2z_{N-2}\prod_{i<j}^{N-1}&\lvert z_{ij}\rvert^{2s_{ij}}\sigma\Bigg\{\prod_{k=2}^{\lfloor N/2 \rfloor}\sum_{m=1}^{k-1}\frac{s_{mk}}{z_{mk}}\prod_{k=\lfloor N/2\rfloor+1}^{N-2}\sum_{n=k+1}^{N-1}\frac{s_{kn}}{z_{kn}}\Bigg\}\nonumber\\&\times\pi\Bigg\{ \prod_{k=2}^{\lfloor N/2 \rfloor}\sum_{m=1}^{k-1}\frac{s_{mk}}{\Bar{z}_{mk}}\prod_{k=\lfloor N/2\rfloor+1}^{N-2}\sum_{n=k+1}^{N-1}\frac{s_{kn}}{\Bar{z}_{kn}}\Bigg\}\,.
	\label{eq: closedEulerSel}
\end{align}
Performing a low-energy expansion of the matrix given in (\ref{eq: closedEulerSel}) gives the following result in terms of the $M$ matrices associated with the open-string (\ref{eq: mmatrix})
\begin{align}
	\mathcal{S}_{\text{tree}}^\mathfrak{m}=S_0\phi^{-1}\left(\sum_{p=0}^{\infty}\,\sum_{i_1,i_2,\ldots,i_p\in 2\NN+1} M_{i_1}\ldots M_{i_p}\sum_{k=0}^p f_{i_1}\ldots f_{i_k}\shuffle f_{i_p}f_{i_{p-1}}\ldots f_{i_{k+1}}\right)\,,
	\label{eq: stree}
\end{align}
where $S_0$ is the field theory limit of the closed-string integral matrix. The $\shuffle$ denotes the shuffle product on the non-commutative words in the $f$-alphabet \cite{Brown:2011ik}. The expression inside the parentheses resembles that of (\ref{eq: ftree}) and is in fact directly related to it via the single-valued map \cite{Stieberger:2013wea,Schlotterer:2018zce,Brown:2019wna} (see \cite{Schnetz:2013hqa,Brown:2013gia,Brown:2018omk} for more details on the single-valued map).

\subsubsection{Four- and five-point examples}
\label{ssec:exmpls}
As this paper is concerned with five-point closed-string amplitudes at one loop, it is helpful to spell out tree-level examples of four- and five-point closed-string interactions. The Mandelstam variables (\ref{eq: mandels}) are not all independent and satisfy constraints arising from momentum conservation, on-shell conditions, and dimensional considerations. As long as the number of external momenta $N$ is less than or equal to the spacetime dimension $D$, we have no additional constraints beyond momentum conservation. As we will work in $D=10$ and consider at most $N=5$, we find from momentum conservation $N(N-3)/2$ independent Mandelstam variables. For $N=4$ this gives the familiar $s:=s_{12}=s_{34},t:=s_{13}=s_{24}, u:=s_{14}=s_{23}$. The other ingredients for $N=4$ are 
\begin{align}
	A_{\rm YM}=A_{\rm YM}(1,2,3,4)\,,\qquad S_0=\frac{\pi su}{t}\,,\qquad M_w=-\frac{1}{w}(s^w+t^w+u^w)\,,
	\label{eq: asm4}
\end{align}
where all quantities are scalar due to the fact that, for $N=4$, the permutation group is trivial. Instead of using the compact motivic $f$-alphabet notation (\ref{eq: stree}), we explicitly write the closed-string low-energy expansion using ordinary zeta values
\begin{align}
	\mathcal{M}_{\text{tree}}^{\rm 4pt}&=A_{\text{YM}}(1,2,3,4)\frac{\pi su}{t}\Big(1+2\zeta_3M_3+2\zeta_5M_5+2\zeta_3^2M_3^2 +2\zeta_7M_7+4\zeta_3\zeta_5M_3M_5+2\zeta_9M_9\nonumber\\&\quad+\tfrac{4}{3}\zeta_3^3M_3^3+2\zeta_5^2M_5^2+4\zeta_3\zeta_7M_3M_7+2\zeta_{11}M_{11}+4\zeta_3^2\zeta_5M_3^2M_5+\dots\Big)\tilde{A}_{\text{YM}}(1,2,3,4)\,.
	\label{eq: 4pttree}
\end{align}
An important feature of this amplitude is the absence of MZVs due to the fact that MZVs arise only as coefficients of commutators of $M$'s.\\
\indent For $N=5$, we find five independent external Mandelstam invariants. Furthermore, there is a nontrivial permutation group of order 2 so that the $A_{\rm YM}$ become two-component vectors and the $S_0, M_{2k+1}$ become $2\times 2$ matrices
\begin{align}
	A_{\rm YM}=\begin{pmatrix}
		A_{\rm YM}(1,2,3,4,5)\\
		A_{\rm YM}(1,3,2,4,5)
	\end{pmatrix}\,,\qquad S_0=\frac{\pi^2}{s_{14}s_{25}s_{35}}\begin{pmatrix}
		\sigma_{11}&\sigma_{12}\\
		\sigma_{21}&\sigma_{22}
	\end{pmatrix}\,,
	\label{eq: 5ptingred}
\end{align}
where $\sigma_{12}=\sigma_{21}=-s_{12}s_{34}s_{13}s_{24}(s_{45}+s_{15})$, $\sigma_{22}=-s_{13}s_{24}(s_{12}s_{23}s_{45}+\rm cyclic(12345))=\sigma_{11}\rvert_{2\leftrightarrow 3}$. The $M$-matrices become more involved, but always enjoy the symmetry $m_{22}=m_{11}\rvert_{2\leftrightarrow 3}$ and $m_{21}=m_{12}\rvert_{2\leftrightarrow 3}$. Explicit expressions for the degree $2k+1$ polynomials in the Mandelstams were calculated in \cite{Schlotterer:2012ny,Oprisa:2005wu,Stieberger:2006bh,Stieberger:2006te,Boels:2013jua} and are stored on the website \cite{MZVWebsite}. With these ingredients, the closed-string tree-level low-energy expansion becomes 
\begin{align}
	\mathcal{M}_{\text{tree}}^{\rm 5pt}&=\begin{pmatrix}
		A_{\rm YM}(1,2,3,4,5)&
		A_{\rm YM}(1,3,2,4,5)
	\end{pmatrix}\frac{\pi^2}{s_{14}s_{25}s_{35}}\begin{pmatrix}
		\sigma_{11}&\sigma_{12}\nonumber\\
		\sigma_{21}&\sigma_{22}
	\end{pmatrix}\Big(1+2\zeta_3M_3+2\zeta_5M_5\nonumber\\&\quad+2\zeta_3^2M_3^2 +2\zeta_7M_7+2\zeta_3\zeta_5\{M_3,M_5\}+2\zeta_9M_9+\tfrac{4}{3}\zeta_3^3M_3^3+2\zeta_3\zeta_7\{M_3,M_7\}+2\zeta_5^2M_5^2\nonumber\\&\quad+2\zeta_{11}M_{11}+2(\tfrac{1}{5}\zeta_{3,3,5}-\tfrac{4}{35}\zeta_2^3\zeta_5+\tfrac{6}{25}\zeta_2^2\zeta_7+9\zeta_2\zeta_9)[M_3,[M_3,M_5]]\nonumber\\&\quad+\zeta_3^2\zeta_5\{M_3,\{M_3,M_5\}\}+\dots\Big)\times\begin{pmatrix}
		\tilde{A}_{\rm YM}(1,2,3,4,5)\\
		\tilde{A}_{\rm YM}(1,3,2,4,5)
	\end{pmatrix}\,,
	\label{eq: 5pttree}
\end{align}
where $[\bullet,\bullet]$ and $\{\bullet,\bullet\}$ denote the commutator and anti-commutator respectively. An interesting observation compared to the $N=4$ amplitude is the appearance of new combinations of $M$ matrices with new coefficients. In later sections, we will see echoes of this feature at genus one. Note that in the above expressions for $N=5$, we have not committed to a basis of Mandelstam variables. For later purposes, it is convenient to introduce a cyclically symmetric set as a kinematic basis denoted $\{s_{12},s_{23},s_{34},s_{45},s_{15}\}$.
%

\subsection{Massless genus-one closed-superstring amplitudes}
\label{sec:oneloop}
%

Having introduced genus-zero closed-superstring amplitudes in the previous section, we advance to genus one. One-loop closed-string amplitudes are expressed as integrals over the moduli space of the punctured torus, which is the genus-one analogue of the tree-level integral (\ref{eq: closedEulerSel}). The punctures represent the positions of the external string states on the toroidal worldsheet. This family of integrals can be split into an integral over the moduli space of the torus without punctures, and an integral over the configuration space of the punctures. We consider the latter integral as the integrand of the former, and it will be referred to as such. As we will see in section \ref{sec:gen1}, the low-energy expansion of the integrand turns out to have a natural description in terms of a type of non-holomorphic modular form\footnote{Note that this modular symmetry that arises from the large diffeomorphisms of the torus acts on completely different objects than the modular symmetry that arises from $S$-duality as treated in section \ref{ssec: sRtypeIIB}.} known as modular graph forms due to their graphical definition \cite{Green:1999pv,Green:2008uj,DHoker:2015gmr,DHoker:2015wxz,DHoker:2015sve,DHoker:2016mwo,DHoker:2016quv,Basu:2016kli,DHoker:2017zhq,Kleinschmidt:2017ege,Gerken:2018zcy,Gerken:2020aju}. We will briefly introduce them and here (see \cite{Gerken:2020xte} for a more extensive review). 

Modular graph forms define a class of non-holomorphic modular forms and are associated to directed labeled simple graphs. Let $\Gamma$ be a graph with each edge $e$ labeled by a pair of integers ($a_e,b_e)$, called the holomorphic and anti-holomorphic labels of the edge $e$ respectively. We assign to any such graph the following series\footnote{This definition does not converge for arbitrary labels $(a_e,b_e)$. We will only deal with convergent instances, and for discussions on divergent MGFs we refer the reader to \cite{Gerken:2020aju}.}
\begin{equation}
	\mathcal{C}_\Gamma(\tau)= \prod_{e\in E_\Gamma}\sum_{p_e\in\Lambda^\prime}\frac{1}{p_e^{a_e}\Bar{p}_e^{b_e}}\prod_{i\in V_\Gamma}\delta\left(\sum_{e'\in E_\Gamma}\Gamma_{ie' }p_{e'}\right)\,,
	\label{eq: mgfdef}
\end{equation}
with $E_\Gamma$ and $V_\Gamma$ the edge and vertex set of $\Gamma$ respectively and $\Gamma_{ie}$ its incidence matrix. The dependence on the complex parameter $\tau=\tau_1+i\tau_2\in\HH$ is included in the sum over the lattice momenta $p_e$ taking values in $\Lambda^\prime=(\mathbb{Z}+\tau\mathbb{Z})\backslash\{0\}$. The $\delta$ denotes the Kronecker delta in the sense $\delta\left(m\tau+n\right) \coloneqq \delta(m)\delta(n)$ with the $\delta$'s on the right-hand side being the usual Kronecker deltas for $m,n\in\mathbb{Z}$. This series defines a non-holomorphic modular form called a modular graph form with modular weights $(|A|,|B|):=(\sum_{e\in E_\Gamma} a_e,\sum_{e\in E_\Gamma} b_e)$. A well-known example of a non-holomorphic modular form of weights $(0,0)$ is the non-holomorphic (real-analytic) Eisenstein series defined by
\begin{equation}
	\EE_s(\tau)=\left(\frac{\tau_2}{\pi}\right)^s\sum_{p\in\Lambda^\prime}\frac{1}{|p|^{2s}}\,,
	\label{eq: nonholEis}
\end{equation}
with $s\in\mathbb{C}$ and $\Re(s)>1$ for convergence. We can represent it as a modular graph form via
\begin{align}
	\left(\frac{\pi}{\tau_2}\right)^s \EE_s(\tau)\,= 
	\tikzset{->-/.style={decoration={
				markings,
				mark=at position .9 with {\arrow{latex}}},postaction={decorate}}}
	\begin{tikzpicture}[baseline={([yshift=-1.5ex]current bounding box.center)}]
		\node (1) at (0,0) {$1$};
		\node (2) at (3.5,0) {$2$};
		\draw[->-] (1) to[bend left=30]
		node[label,fill=white]{$(s,0)$} (2);
		\draw[->-] (1) to[bend right=30]
		node[label,fill=white]{$(0,s)$} (2);
	\end{tikzpicture}\,,
	\label{eq: dihedral}
\end{align}
where the equality is an abuse of notation: it is actually equal to the lattice sum associated with the graph, not the graph itself. Note that the labeling of the vertices is arbitrary.\\
\indent A more general class of MGFs that plays a role in the calculation of genus-one closed-string amplitudes is the dihedral modular graph form, which we denote with
\begin{align}
	\tikzset{->-/.style={decoration={
				markings,
				mark=at position .9 with {\arrow{latex}}},postaction={decorate}}}
	\cforml{a_1&\dots&a_r\\b_1&\dots&b_r}:=\sum_{p_1,\dots,p_r\in(\Lambda^\prime)^r}\frac{\delta(p_1+\dots+p_r)}{p_1^{a_1}\Bar{p}_1^{b_1}\dots p_r^{a_r}\Bar{p}_r^{b_r}} =\begin{tikzpicture}[baseline={([yshift=-2ex]current bounding box.center)}]
		\node (1) at (0,0) {$1$};
		\node (2) at (3.5,0) {$2$};
		\draw[->-] (1) to[bend left=50]
		node[label,fill=white]{$(a_{1},b_{1})$} (2);
		\draw[->-] (1) to[bend left=20]
		node[label,fill=white]{$(a_{2},b_{2})$} (2);
		\node at (1.75,-0.1) {$\vdots$};
		\draw[->-] (1) to[bend right=40]
		node[label,fill=white]{$(a_{r},b_{r})$} (2);
	\end{tikzpicture}\,.
	\label{eq: dihedralC} 
\end{align}
Similar notation exists for more complicated graph topologies. As we will mainly employ the iterated integral formalism, we refrain from spelling out all the notational details of higher-point topologies here and refer to \cite{Gerken:2020aju} instead.

Although modular graph forms appear naturally in one-loop string integrands, their intricate relations \cite{DHoker:2015gmr,DHoker:2015sve,DHoker:2016mwo,DHoker:2016quv, Gerken:2018zcy,Basu:2016kli,Gerken:2020aju,Kleinschmidt:2017ege} (which are generally unknown beyond total modular weight greater than 12) complicate their presentation and hide their algebraic properties. Furthermore, as MGFs represent convoluted lattice-sums whose integrals over the moduli space of the torus are not easily evaluated, they are not the ideal objects to consider at high orders in the low-energy expansion.

Recently, an alternative representation of MGFs has been put forward through a class of non-holomorphic modular forms known as equivariant iterated Eisenstein integrals (EIEIs) \cite{Brown:2017qwo,Brown:2017qwo2,Broedel:2018izr,Gerken:2019cxz,Gerken:2020yii,Gerken:2018jrq,Dorigoni:2022npe,Dorigoni:2024oft,Claasen:2025vcd}. This alternative basis lends itself for a much neater organization of contributions to the integrand and is easier to handle when integrating over the moduli space of the torus. To construct these iterated integrals of holomorphic Eisenstein series, we require a suitable set of integration kernels. For our purposes, a particularly convenient choice is the following \cite{Dorigoni:2022npe}
\begin{align}
	\omplus{j}{k}{\tau,\tau_1}&=\frac{d\tau_1}{2\pi i}\left(\frac{\tau-\tau_1}{4y}\right)^{k-2-j}\left(\Bar{\tau}-\tau_1\right)^j\GG_k(\tau_1)\,,\nonumber\\\omminus{j}{k}{\tau,\tau_1}&=-\frac{d\Bar{\tau}_1}{2\pi i}\left(\frac{\tau-\Bar{\tau}_1}{4y}\right)^{k-2-j}\left(\Bar{\tau}-\Bar{\tau}_1\right)^j\overline{\GG_k(\tau_1)}\,,
	\label{eq: kernels}
\end{align}
with $k\geq 4$ even, $0\leq j\leq k-2$ and $y=\pi \tau_2$. Please note that in these definitions and other similar instances, $\tau_1$ is not the real part of $\tau$, but merely an auxiliary integration variable. $\GG_k(\tau)$ is the holomorphic Eisenstein series
\begin{align}
	\GG_k(\tau) = \sum_{p\in\Lambda^\prime}\frac{1}{p^k} \quad\quad k\geq4\, ,
\end{align}
and it is a holomorphic modular form of weight $k$. For $k=2$ the sum is conditionally convergent. One way to define it is to use the following summation prescription
\begin{align}
	\hat\GG_2=\sum_{n\neq0}\frac{1}{n^2}+\sum_{m\neq0}\sum_{n\in\mathbb{Z}}\frac{1}{(m\tau+n)^2} - \frac{\pi}{2}\, ,
\end{align}
which is called the Eisenstein summation. This is a modular function of weight two but it is not holomorphic anymore. Even though $\hat\GG_2$ might appear in the middle stages of the calculations (see \eqref{eq: exHSR}), for the EIEIs we always use $\GG_k$ with $k\geq4$. The kernels (\ref{eq: kernels}) are designed to transform as modular forms with modular weights $(0,k-2j-2)$. Even though the kernels depend non-holomorphically on $\tau$, their corresponding iterated integrals
\begin{align}
	\betaplustau{j_1&j_2&\cdots&j_\ell\\k_1&k_2&\cdots&k_\ell}&=\int_\tau^{i\infty}\omplus{j_\ell}{k_\ell}{\tau,\tau_\ell}\dots\int_{\tau_3}^{i\infty}\omplus{j_2}{k_2}{\tau,\tau_2}\int_{\tau_2}^{i\infty}\omplus{j_1}{k_1}{\tau,\tau_1}\,,\nonumber\\\betaminustau{j_1&j_2&\cdots&j_\ell\\k_1&k_2&\cdots&k_\ell}&=\int_{\Bar{\tau}}^{-i\infty}\omminus{j_\ell}{k_\ell}{\tau,\tau_\ell}\dots\int_{\Bar{\tau}_3}^{-i\infty}\omminus{j_2}{k_2}{\tau,\tau_2}\int_{\Bar{\tau}_2}^{-i\infty}\omminus{j_1}{k_1}{\tau,\tau_1}\,,
	\label{eq: betapm}
\end{align}
are homotopy invariant \cite{Dorigoni:2022npe}. To establish some nomenclature, we call the sum $k_1+\dots+k_\ell$ the degree of the iterated integral and $\ell$ the depth.\\
\indent The vector space $\mathscr{M}_k$ of weight-$k$ holomorphic modular forms contains, in addition to the $\GG_k(\tau)$ appearing in (\ref{eq: kernels}), also holomorphic cusp forms $\Delta_k(\tau)$. In fact, the space $M_k$ decomposes as $\mathscr E_k\oplus\mathscr{S}_k$ with $\mathscr{E}_k$ the space of Eisenstein series of weight $k$ and $\mathscr{S}_k$ the space of cusp forms of weight $k$\footnote{The dimension of $\mathscr{E}_k$ is one for all $k$, whereas ${\rm dim}\, \mathscr{S}_k$ can be greater than one.}. Similarly to the construction of iterated integrals of $\GG_k$ in (\ref{eq: betapm}), we can construct iterated integrals of cusp forms where we replace $\GG_k$ by a cusp form $\Delta_k\in\mathscr{S}_k$. We denote the cuspidal kernels with $\ompm{j}{\Delta_k}{\tau,\tau_1}$ and replace $k_i\mapsto (2\pi i)^{k_i}\Delta_{k_i}$ in (\ref{eq: betapm}) to define iterated integrals of cusp forms. \\
\indent Although the kernels (\ref{eq: kernels}) are modular, their iterated integrals (\ref{eq: betapm}) are not. Nevertheless, it is possible to find modular completions of them \cite{Brown:mmv,Brown:2017qwo,Brown:2017qwo2,Dorigoni:2024oft}. For iterated Eisenstein integrals, these completions have been spelled out for depth $\leq 3$ and degree $\leq 20$ in \cite{Dorigoni:2022npe,Dorigoni:2024oft,Brown:2017qwo,Brown:2017qwo2,Drewitt:2021}. Given an iterated integral of the form (\ref{eq: betapm}), we denote its modular completion with $\beta^{\mathrm{eqv}}$. It is modular covariant and transforms as
\begin{align}
	\betaeqvx{j_1&j_2&\dots&j_\ell\\k_1&k_2&\dots&k_\ell}{\tfrac{a\tau +b}{c\tau +d}}=\left(\prod_{i=1}^\ell (c\Bar{\tau}+d)^{k_i-2j_i-2}\right)\betaeqvtau{j_1&j_2&\dots&j_\ell\\k_1&k_2&\dots&k_\ell}\,,
	\label{eq: betacovariant}
\end{align}
with modular weights $(0,\sum_{i=1}^\ell k_i-2j_i-2)$. To give an idea of what such completions look like, the following combination trivializes the $T$- and $S$-cocycles (where $T$ and $S$ are understood as the two generators of $\SL(2,\mathbb Z)$) of the $\beta_+$ 
\begin{align}
	\betaeqvx{j\\k}{\tau}:=\betaplustau{j\\k}+\betaminustau{j\\k}-\frac{2\zeta_{k-1}}{(k-1)(4y)^{k-2-j}}\,,
	\label{eq: d1beqv}
\end{align}
so that it transforms as a modular form of weights $(0,k-2-2j)$. The ingredients for completions at higher depths were shown to involve iterated integrals of $\GG_k$ and $\Delta_k$ of depth $\leq 3$ together with multiple modular values such as the L-values of cusp forms and multiple zeta values \cite{Brown:mmv,Dorigoni:2024oft}. A particular example of a modular iterated integral of depth $\ell=2$ is
\begin{align} \betaeqv{8&2\\10&4}&=\betaplus{8&2\\10&4}+\betaminus{2&8\\4&10}+\betaplus{2\\4}\betaminus{8\\2}-\frac{2}{9}\betaplus{2\\4}\zeta_9-\frac{2}{3}\betaminus{8\\10}\zeta_3\nonumber\\&+\frac{2 i \pi^9\Bar{\tau}^9 \zeta_3}{2525985}-\frac{i\pi^3\Bar{\tau}^3\zeta_9}{1215}+\frac{2\zeta_3\zeta_9}{27}+\frac{\Lambda(\Delta_{12},12)}{122472000\Lambda(\Delta_{12},10)}\left(\betaplus{10\\\Delta_{12}}-\betaminus{10\\\Delta_{12}}\right)\,,
	\label{eq: betaeqvexample}
\end{align}
which is a modular form of weights $(0,-10)$. Here, $\Lambda(\Delta_k,t)$ is the completed L-function for a cusp form $\Delta_k$ \cite{Dorigoni:2021ngn}. Note that we omitted the explicit $\tau$ dependence of the iterated integrals to prevent cluttering of notation. For more examples, see \cite{Dorigoni:2022npe,Dorigoni:2024oft}.\\
\indent As we shall see in the next section, the conversion algorithm that translates MGFs to iterated Eisenstein integrals never deals with cusp forms and any transcendental periods other than MZVs. This implies that we will never encounter iterated integrals of cusp forms when we restrict ourselves to the space spanned by MGFs. In the generating series of all equivariant iterated Eisenstein integrals, this dropout of cusp forms is realized through certain relations on the non-commutative bookkeeping variables. This algebra is known as Tsunogai's derivation algebra, with relations called Tsunogai or Pollack relations \cite{Pollack,Tsunogai,Dorigoni:2022npe,Brown:2017qwo2,Gerken:2019cxz,Gerken:2020yii,Dorigoni:2021ngn}. The generating series of MGFs in \cite{Dorigoni:2022npe,Dorigoni:2024oft} conjecturally realizes matrix representations of these variables $\epsilon_k$ so that the holomorphic cusp forms cancel \cite{Gerken:2019cxz,Gerken:2020yii}. \\
\indent A crucial property of modular iterated Eisenstein integrals relevant to the conversion algorithm is their differential equation with respect to $\tau$ \cite{Dorigoni:2022npe,Dorigoni:2024oft} given by
\begin{align}
	\pi\nabla_0\betaeqvtau{j_1&j_2&\cdots&j_\ell\\k_1&k_2&\cdots&k_\ell}&=-\frac{1}{4}\sum_{i=1}^\ell(k_i-j_i-2)\betaeqv{j_1&\cdots&j_i+1&\cdots&j_\ell\\k1&\cdots&k_i&\cdots&k_\ell}\nonumber\\&\quad+\frac{1}{4}\delta_{j_\ell,k_\ell-2}(2 i\tau_2)^{k_\ell}\GG_{k_\ell}(\tau)\betaeqvtau{j_1&j_2&\cdots&j_{\ell-1}\\k_1&k_2&\cdots&k_{\ell-1}}+\dots\,,
	\label{eq: betaeqvdiffeqn}
\end{align}
where $\nabla_0:=2i\tau_2^2\partial_{\tau}$. The ellipsis involves (iterated integrals of) cusp forms. As cusp forms are absent in the world of MGFs, we can ignore those terms for our purposes. Given a specific right-hand side of the differential equation, one can find a solution within the space of $\beta^{\mathrm{eqv}}$ under specific conditions that are always satisfied in the world of MGFs, and we do not study further in this work. The modularity of both sides of (\ref{eq: betaeqvdiffeqn}) generally provides the boundary conditions and fixes most integration constants. The remaining integration constants can be fixed by requiring consistency of the system of differential equations for all $\beta^{\text{eqv}}$, which still leaves some constants that can be fixed by gauge-fixing the generating series of EIEIs \cite{Dorigoni:2024oft}.\\
\indent One of the benefits of using this language in the context of genus-one string amplitudes is the fact that there are no linear relations among the $\beta^\eqv$ with different entries $j_i,k_i$ \cite{Matthes_2017}. This guarantees a canonical basis for MGFs. 

\subsubsection{Conversion algorithm}
\label{ssec:conv}
We now have two languages at our disposal that define certain classes of non-holomorphic modular forms: the modular graph forms that naturally arise from string theory, and the equivariant iterated Eisenstein integrals. As the mathematical structure of the latter function space is much more transparent, it would be beneficial to be able to systematically translate MGFs to EIEIs. This has recently been achieved through a conversion algorithm \cite{Claasen:2025vcd} based on the sieve algorithm for MGFs \cite{DHoker:2016mwo,Gerken:2020xte,Hidding:2022zzz} and has been implemented in a Mathematica package \textsc{MGFtoBeqv} (see \url{https://github.com/emieltcc/MGFtoBeqv}). \\
\indent The main idea of the algorithm is to first ``decompose'' a modular graph form in terms of holomorphic Eisenstein series $\GG_k(\tau)$ by repeatedly acting with a differential operator on it. Throughout this process, specific obstructions can occur. We need to remove these obstructions using Fay identities, which produce the desired factors of $\GG_k(\tau)$. This process defines a family of $\GG_k(\tau)$ scattered around the different levels in the iterative differential process. Using this decomposed representation of the MGF, we iteratively integrate this system back up. This gives the iterated Eisenstein integral representation of the MGF. The main ingredients of this procedure will be reviewed in the following, along with an example.\\
\indent The differential operators that we need are the Maass operators introduced in section \ref{ssec: sRtypeIIB}. Their action on modular graph forms is given by shifting the exponents of (anti-)holomorphic momentum by (minus) one
\begin{align}
	\nabla^{(\lvert A\rvert)}\mathcal{C}_\Gamma=\sum_{e\in E_\Gamma}a_e\mathcal{C}_{\Gamma_{(a_e,b_e)\mapsto (a_e+1,b_e-1)}}\,,\quad\overline{\nabla}^{(\vert B\rvert)}\mathcal{C}_{\Gamma}=\sum_{e\in E_\Gamma}b_e\mathcal{C}_{\Gamma_{(a_e,b_e)\mapsto (a_e-1,b_e+1)}}\,.
	\label{eq:maassmgfs}
\end{align}
The conversion algorithm is best suited for a different convention of MGFs that is amenable to its description in terms of EIEIs. As EIEIs are defined to have purely anti-holomorphic weight and MGFs in general can have both, it is natural to remove the holomorphic weights of MGFs by multiplying with an appropriate power of $\tau_2$ (which has modular weights $(-1,-1)$). This results in redefining MGFs of total weights $(\lvert A\rvert,\lvert B\rvert)$ to MGFs of weights $(0,\lvert B\rvert-\lvert A\rvert)$. Furthermore, for convenience, we also scale with factors of $\pi$ 
\begin{align}
	\mathcal{C}_\Gamma^+(\tau):=\frac{\tau_2^{\lvert A\rvert}}{\pi^{(\lvert A\rvert+\lvert B\rvert)/2}}\mathcal{C}_\Gamma(\tau)\,.
	\label{eq:cplus}
\end{align}
The holomorphic differential operator $\nabla^{(0)}=2i\tau_2\partial_\tau$ takes us out of the space of MGFs of purely anti-holomorphic weight. Hence, we use the variant $\nabla_0$ that keeps us within the space of MGFs of purely anti-holomorphic weight.

The obstructions which we alluded to at the beginning of this section are called closed holomorphic subgraphs \cite{DHoker:2016mwo,Gerken:2018zcy}. These are defined as closed subgraphs of an MGF whose edges have zero anti-holomorphic weight. They naturally arise upon the action of the Maass operators, which can be inferred from the way they change the edge labels of an MGF in (\ref{eq:maassmgfs}). Closed holomorphic subgraphs define obstructions in the sense that acting with another Maass operator on an MGF with a closed holomorphic subgraph produces edges with negative edge labels within this subgraph. As explained in \cite{Claasen:2025vcd}, such edges make it impossible to recover the integration kernels $\GG_k(\tau)$. To remove these obstructions, we can employ special cases of (genus-one) Fay identities \cite{Fay}, which in the context of MGFs define the technique of holomorphic subgraph reduction (HSR) \cite{DHoker:2016mwo,Gerken:2018zcy}. HSR works recursively: it translates an $n$-point closed holomorphic subgraph to $(n-1)$-point closed holomorphic subgraphs, plus graphs of lower loop order. By repeated application of this technique, any closed holomorphic subgraph can be completely removed and produces other MGFs free of obstructions that may multiply $\GG_k(\tau)$ for some $k\geq 4$. We emphasize that this procedure is extremely general and can be used to remove all obstructions in the algorithm for any number of points and weights. For example, given the specific dihedral MGF below with a closed holomorphic subgraph, one can reduce it as follows \cite{DHoker:2016mwo}
\begin{align}
	\cforml{1& 2 & 2\\1 & 0 & 0}=-6 \cforml{5 & 0 \\1 & 0}+ 2\cforml{3 & 0\\1 & 0}\hat{{\rm G}}_2+\frac{2\pi}{\tau_2}{\rm G}_4\,.
	\label{eq: exHSR}
\end{align}

To summarize the algorithm and get a high-level idea of what is happening to a particular (linear combination of) graph(s), we find the following visualization helpful \cite{Claasen:2025vcd}. A linear combination of MGFs is visualized by a solid black node. The action of $\pi\nabla_0$ together with the removal of obstructions is represented by an edge for every $\GG_k(\tau)$ series created, plus one edge for the terms that do not multiply any $\GG_k$ (if there are any). The edges connect from the original node to new nodes that represent the produced MGFs that are the coefficients of the $\GG_k$ next to the node
\begin{align}
	\begin{tikzpicture}[baseline=(zero.base)]
		\draw (0,1) to (-2,0);
		\draw (0,1) to (-1,0);
		\draw (0,1) to (2,0);
		\fill (0,1) circle (2pt);
		\fill (-2,0) circle (2pt);
		\fill (-1,0) circle (2pt);
		\fill (2,0) circle (2pt);
		\node[right] at (-0.95,0) {$\GG_{k_1}$};
		\node[right] at (2,0) {$\GG_{k_n}$};
		\node[right] at (0,1/4) (zero) {$\boldsymbol{\cdots}$};
		\draw[->,-Stealth](-3,0.9)--(-3,0.1);
		\draw(-3.5,0.5)node{$\pi\nabla_0$};
	\end{tikzpicture}\,.
	\label{eq: singlestep}
\end{align}
This process can be repeated for every solid black node that appears at the bottom, creating a tree structure of solid nodes decorated with $\GG_k$. This process is guaranteed to terminate after $\lvert B\rvert$ iterations \cite{Claasen:2025vcd}. To visualize the termination, we depict the last/lowest nodes (that represent rational numbers) as open nodes.\\
\indent To illustrate, we consider the dihedral MGF (\ref{eq: dihedralC}) with four lines, all with weights $(1,1)$. This MGF appears in the calculation of both the four- and five-point closed-string integrands (also known as the three-loop banana graph). We can completely decompose it in terms of holomorphic Eisenstein series iterating the procedure described above four times. The result can be visualized as
\begin{align}
	\begin{tikzpicture}[baseline={([yshift=-0.5ex]current bounding box.center)}]
		\draw[] (1,1) to (1,0); 
		\draw[] (1,0) to (0,-1);
		\draw[] (1,0) to (2,-1);
		\draw[] (0,-1) to (-1,-2);
		\draw[] (0,-1) to (1,-2);
		\draw[] (2,-1) to (3,-2);
		\draw[] (-1,-2) to (-1,-3);
		\draw[] (1,-2) to (1,-3);
		\draw[] (3,-2) to (3,-3);
		\fill (-1,-2) circle (2pt);
		\fill (1,-2) circle (2pt);
		\fill (3,-2) circle (2pt);
		\fill (1,0) circle (2pt);
		\fill (0,-1) circle (2pt);
		\fill[color=orange] (2,-1) circle (2pt);
		\draw[fill=white] (-1,-3) circle[radius=2pt];
		\draw[fill=white] (1,-3) circle[radius=2pt];
		\draw[fill=white] (3,-3) circle[radius=2pt];
		\node[right] at (-1,-3) {$\GG_8$};   
		\node[right] at (1,-3) {$\GG_4$}; 
		\node[right] at (3,-3) {$\GG_4$};
		\node[right] at (2,-1) {$\GG_4$};
		\node[right] at (1,-2) {$\GG_4$};
		\node[fill=white] at (1,1) {\textcolor{orange}{$\cformp{1\,1\,1\,1\\1\,1\,1\,1}$}};    
		\node[fill=white] at (-2,1) {$\#$ of $\pi\nabla_0$};
		\node[fill=white] at (-2,0) {$1$};
		\node[fill=white] at (-2,-1) {$2$};
		\node[fill=white] at (-2,-2) {$3$};
		\node[fill=white] at (-2,-3) {$4$};
	\end{tikzpicture}\,,
	\label{eq: othertrees}
\end{align}
where the orange color is explained in the last paragraph of this section. The superscript of the dihedral MGF refers to the fact that we changed conventions using (\ref{eq:cplus}) compared to (\ref{eq: dihedralC}).\\
\indent Now we are in a position to integrate the system back up and construct the iterated Eisenstein integral representation of the MGF. We do this by starting at the bottom of the tree structure and take each branch of the tree as a right-hand side of the differential equation (\ref{eq: betaeqvdiffeqn}). We can solve this differential equation, which is just a linear algebra problem, to move one step up the tree. Arriving at the next node, we are instructed to perform one of the following actions:
\begin{itemize}
	\item If a $\GG_k$ appears, we multiply the solution of the equation by $\GG_k$.
	\item If two branches combine, we add the solutions.
	\item If none of the aforementioned occurs, we do nothing.
\end{itemize}
We now view these results again as separate right-hand sides of the differential equation (\ref{eq: betaeqvdiffeqn}) and solve it. We repeat this procedure until we arrive at the top of the tree, which gives
\begin{align}
	\cformp{1\,1\,1\,1\\1\,1\,1\,1}=-504 \betaeqv{3\\8} + 216 \betaeqv{1&1\\4&4}-432 \betaeqv{2&0\\4&4}\,.
	\label{eq:convexamp}
\end{align}

In solving the differential equations, the boundary conditions are generally supplied by modularity of both sides. This is sufficient to fix the integration constants, except when the solution is modular invariant. In this case, we have the freedom to add any constant to the solution. To solve the system uniquely, we have to supply the boundary condition from the constant in the Laurent polynomial of the original linear combination of MGFs that appeared at that vertex. We colored the vertices where this happens orange to keep track of them. For more details and examples, see \cite{Claasen:2025vcd}.

\section{Massless genus-one five-point amplitudes}
\label{sec:gen1}
In this section, we calculate one-loop contributions to the low-energy effective action of five-point string amplitudes up to $D^{12}R^5$ and $D^{14}\phi R^4$ in both the ${\rm U}(1)$ conserving and violating sectors respectively. As the four-point result plays a large role in the five-point one, we first go through the four-point calculation in section \ref{ssec:4ptrev}. In section \ref{ssec:gen5pt}, we write the general formula of the five-point amplitude for any sector, where in sections \ref{ssec: conserv} and \ref{ssec:viol} we focus on the ${\rm U}(1)$-conserving and violating sectors respectively. These sectors turn out to be closely related, as was explained in \cite{Green:2013bza}. We show that this relation continues to hold for the new results \ref{ssec: sdual}.

\subsection{Four-graviton warm-up}
\label{ssec:4ptrev}
The machinery described in section \ref{sec:oneloop} proved fruitful in pushing the calculation of the one-loop four-graviton amplitude in type II string theory in flat 10D spacetime up to $D^{14}R^4$ in the effective action \cite{Claasen:2024ssh}. In this section, we review this calculation, which serves both as a step-up and as an important ingredient in the five-point calculation to be discussed in the rest of this section. The four-graviton amplitude in ten-dimensional flat spacetime in Type II string theory at one loop can be written as \cite{Green:2013bza}
\begin{align}
	\mathcal{M}^{\rm 4pt}=\sum_{h=0}^\infty g_s^{2h-2}A_{\rm YM}\mathcal{S}_{h}^{\rm 4pt}\tilde{A}_{\rm YM}\,,
	\label{eq: altamp}
\end{align}
up to overall normalization. For $h=0$, the kernel can be identified with the $N=4$ variant of $\mathcal{S}_{\text{tree}}$ from (\ref{eq: clstring}). After factoring out the field theory limit $S_0$, the coefficient functions ${\mathcal{S}}^{\rm 4pt}_{h}(s_{ij})$ are invariant under permutations of the $s_{ij}$ variables, implying that their momentum dependence can be captured with the symmetric polynomials $M_k$ from (\ref{eq: asm4}).\\
\indent At tree level, the coefficient function can be compactly written as \cite{Green:1999pv}
\begin{equation}
	{\mathcal{S}}^{\rm 4pt}_{\rm tree}(s_{ij})=S_0\exp\left\{\sum_{m=1}^\infty 2\zeta_{2m+1}M_{2m+1}\right\} \, ,
	\label{eq: tree-sigmas}
\end{equation}
which agrees with (\ref{eq: 4pttree}). At genus one, such a closed formula for ${\mathcal{S}}^{\rm 4pt}_{\rm 1-loop}(s_{ij})$ also exists \cite{Baccianti:2025whd}. However, it is currently unknown how to meaningfully expand those results for low energies, hence making the low-energy expansion inaccessible through this method. Instead, the low-energy expansion of ${\mathcal{S}}^{\rm 4pt}_{\rm 1-loop}(s_{ij})$ can be calculated by expanding the string amplitude integrand for small kinematic variables before integration, so that the associated coefficient functions expand accordingly. The contribution of each term to the low-energy expansion can then be calculated separately by explicit integration.

In order to determine ${\mathcal{S}}^{\rm 4pt}_{\rm 1-loop}(s_{ij})$, we need to perform an integral over the moduli space of the toroidal worldsheet with four punctures representing the external string states. We can split this integral over the moduli space in to a product of integrals: one over the moduli space of the torus, and the others over the configuration space of the punctures.

The torus can be parametrized as the quotient $\Sigma_\tau=\CC/\Lambda$, where $\Lambda = \mathbb{Z} + \tau \mathbb{Z}$ is a lattice determined by the complex structure modulus $\tau = \tau_1 + i\tau_2$. The moduli space $\mathcal{F}$ of the torus is given by $\mathcal{F} = \mathbb{H}/\text{PSL}(2,\mathbb{Z})$. Here, $\HH$ is the complex upper-half plane, and $\text{PSL}(2,\ZZ)$ is the modular group. To provide a concrete geometric representation of this moduli space, one must select a fundamental domain within $\HH$, which we do as follows
\begin{align}
	\mathcal{F}&=\{\tau\in\mathbb{H}\mid-\tfrac{1}{2}\leq \Re(\tau)\leq 0,|\tau|^2\geq 1\}\cup\, \{\tau\in\mathbb{H}\mid 0\leq \Re(\tau)<\tfrac{1}{2},|\tau|^2>1\} \, .
	\label{eq: funddom}
\end{align}
The splitting of the moduli space integrations results in the amplitude \cite{Green:1981yb}
\begin{equation}
	{\mathcal{S}}^{\rm 4pt}_{\rm 1-loop}(s_{ij})=2\pi S_0M_3\int_{\mathcal{F}}\frac{d^2\tau}{\tau_2^2}\mathcal{B}(s_{ij}|\tau)\, .
	\label{eq: g1amplitude}
\end{equation}
The integrand is given by the configuration-space integrals of the punctures (with one fixed by translation invariance) 
\begin{equation}
	\mathcal{B}(s_{ij}|\tau)=\prod_{k=2}^4\int_{\Sigma_\tau}\frac{d^2z_k}{\tau_2}\KN_4\,,\quad \KN_n=\exp\left\{-\color{black}{\sum_{1\leq i<j\leq n} s_{ij}G_{ij}(\tau)}\right\}\,,
	\label{eq: integrandB}
\end{equation}
where we used shorthand notation $G_{ij}(\tau):=G(z_i-z_j|\tau)$ for the Green function on the torus. The exponential is often called the Koba-Nielsen factor, hence the acronym $\KN$. As functions on the torus are doubly-periodic, the Green function admits a double Fourier expansion 
\begin{equation}
	G(z,\tau)=\frac{\tau_2}{\pi}\sum_p^\prime \frac{e^{2\pi i\langle p,z \rangle}}{|p|^2} \, ,
	\label{eq: greensfourier}
\end{equation}
with $p=m\tau+n$ for $m,n\in\mathbb{Z}$. $z$ denotes the coordinate of a puncture on the worldsheet and $\langle p,z\rangle=\tfrac{p\Bar{z}-\Bar{p}z}{2i\tau_2}$. The prime on the summation means $(m,n)\neq (0,0)$.

To perform the low-energy expansion, we Taylor expand the exponential in (\ref{eq: integrandB}) for small values of momenta. However, there is one caveat to this. Namely, in the degeneration limit of the torus one recovers the field theory amplitude, which we know to have non-analytic contributions. Thus, we can only expand the exponential in a region away from the degeneration limit $\tau\rightarrow i\infty$. To this end, we partition the fundamental domain into two regions \cite{Green:1999pv}
\begin{align}
	\mathcal{F}=\mathcal{F}_L\cup \mathcal{F}_R\,,\quad \mathcal{F}_L=\mathcal{F}\cap\{\tau_2\leq L\}\,\,\text{and}\,\, \mathcal{F}_R=\mathcal{F}\cap\{\tau_2>L\}\,,
	\label{eq:split}
\end{align}
where $L$ is an auxiliary cutoff that drops out from the final results. The amplitude (\ref{eq: g1amplitude}) decomposes under the partitioning of $\mathcal{F}$ as
\begin{equation}
	{\mathcal{S}}^{\rm 4pt}_{\rm 1-loop}(s_{ij})={\mathcal{S}}^{\rm 4pt}_{L}(L;s_{ij})+{\mathcal{S}}^{\rm 4pt}_{R}(L;s_{ij})\, ,
	\label{eq: ampsplit}
\end{equation}
where the subscripts of $\mathcal{S}_L$ and $\mathcal{S}_R$ imply that we only integrate $\mathcal{B}(s_{ij}|\tau)$ over $\mathcal{F}_L$ and $\mathcal{F}_R$ respectively. The $\mathcal{S}_R^{\rm 4pt}(L;s_{ij})$ has been calculated explicitly in \cite{DHoker:2019blr}. The remaining integral that we thus consider is
\begin{align}
	\mathcal{S}^{\rm 4pt}_L(L;s_{ij})=2\pi S_0 M_3\int_{\mathcal{F}_L}\frac{d^2\tau}{\tau_2^2}\mathcal{B}(s_{ij}|\tau)\,.
	\label{eq: ampLow}
\end{align}
As the degeneration limit of the torus lies in $\mathcal{F}_R$, we can safely perform a low-energy expansion of this integrand. \\
\indent After expanding the Koba-Nielsen factor in (\ref{eq: integrandB}), we can perform the integrals over the punctures using the Fourier expansion of the Green function (\ref{eq: greensfourier}). After integration, the resulting functions lie exactly in the space of modular graph forms (or actually modular graph functions as they are required to be invariant) defined in section \ref{sec:oneloop}. The process of expanding the Koba-Nielsen factor and integrating over the punctures is completely algorithmic up to any order, but computing the remaining integral over the fundamental domain as in (\ref{eq: g1amplitude}) is a more difficult task. Moreover, there is no canonical basis for these MGFs and there exist many unknown relations at expansion orders higher than 6. To avoid these issues, we use the algorithm explained in section \ref{ssec:conv} to translate the integrand to the function space defined by the equivariant iterated Eisenstein integrals. 

We find\footnote{The result is known up to order $\ap^{11}$ \cite{Claasen:2025vcd}, but it will not be relevant to the five-point computation, hence it is omitted.} 
\begin{align}
	\mathcal{S}^{\rm 4pt}_{L}&(L;s_{ij})=S_0\Big(\Xi_3M_3+\Xi_5M_5+\Xi_{3,3}M_3^2+\Xi_7M_7+\Xi_{\{3,5\}}\{M_3,M_5\} \nonumber\\&+\Xi_9M_9+\Xi_{3,3,3}M_3^3+\beta\, \Xi_{\{3,7\}}\{M_3,M_7\}+ (1-\beta)\Xi_{\{5,5\}}\{M_5,M_5\}+\mathcal{O}(\alpha'^{11})\Big)\,,
	\label{eq: s1loop}
\end{align}
where the coefficients are integrals of linear combinations of iterated Eisenstein integrals. Recall that $M_k$ comprises matrices with entries that are degree $k$ polynomials in $s_{ij}$ and are hence of order $\ap^k$. As it is impossible to distinguish $\{M_3,M_7\}$ from $\{M_5,M_5\}$ at four points since they are equal (which means $\Xi_{\{3,7\}}=\Xi_{\{5,5\}}$), we have a one-parameter freedom in the presentation of the amplitude, captured by the parameter $\beta$. This degeneracy is lifted in the five-point case, but as we will see in section \ref{sec:gen1}, the one-parameter freedom will remain in an interesting manner. The coefficients $\Xi_{\bullet}$ are given by the $\ap$-expansion of (\ref{eq: ampLow}).
Using the results of \cite{Doroudiani:2023bfw}, these integrals can be evaluated and result in
\begin{align}
	&\Xi_3\approx\frac{2\pi^2}{3}\,,\nonumber\\
	&\Xi_5\approx0\,,\nonumber\\
	&\Xi_{3,3}\approx\frac{2\pi^2\zeta_3}{3}\,,\nonumber\\
	&\Xi_{7}\approx\frac{16\pi^2\zeta_3}{45}\bigg(\log(4\pi L)-\frac{25}{12}+\bigg[\gamma_E-\frac{\zeta'_{-3}}{\zeta_{-3}}-\frac{\zeta'_{3}}{\zeta_3}\bigg]\bigg)\,,\nonumber\\
	&\Xi_{\{3,5\}}\approx\frac{29\pi^2\zeta_5}{90}\,,\nonumber\\
	&\Xi_9\approx\frac{88\pi^2\zeta_5}{315}\bigg(\log(4\pi L)+\frac{4831}{1980}+\bigg[\gamma_E-\frac{\zeta'_{5}}{\zeta_5}+\frac{2}{33}\frac{\zeta'_{-5}}{\zeta_{-5}}-\frac{70}{33}\frac{\zeta'_{-3}}{\zeta_{-3}}+\frac{35}{33}\frac{\zeta'_{-1}}{\zeta_{-1}}\bigg] \bigg)\,,\nonumber\\
	&\Xi_{3,3,3}\approx\frac{\pi^2\zeta_3^2}{3}+\frac{58\pi^2\zeta_5}{189}\bigg(\log(4\pi L)-\frac{3028}{1305}+\bigg[\gamma_E-\frac{\zeta'_{5}}{\zeta_5}-\frac{293}{174}\frac{\zeta'_{-5}}{\zeta_{-5}}+\frac{119}{87}\frac{\zeta'_{-3}}{\zeta_{-3}}-\frac{119}{174}\frac{\zeta'_{-1}}{\zeta_{-1}}\bigg]\bigg)\,,\nonumber\\
	&\Xi_{\{3,7\}}\approx -\frac{163\pi^2\zeta_7}{7560}+\frac{86\pi^2\zeta_3^2}{315}\bigg(\log(4\pi L)-\frac{205}{86}+\bigg[\frac{58}{43}\gamma_E-\frac{15}{43}\frac{\zeta'_{-5}}{\zeta_{-5}}-\frac{28}{43}\frac{\zeta'_{-3}}{\zeta_{-3}}-\frac{58}{43}\frac{\zeta'_3}{\zeta_3}\bigg]\bigg)\,,
	\label{eq: xisint4pt}
\end{align}
where the $\approx$ sign means that of all $L$-dependent terms, we only keep the logarithmic one (all $L$-dependent terms cancel in the full amplitude) and omit writing the power-law terms with positive and negative exponents.\\
\indent Note that we purposefully write the results only in terms of logarithmic derivatives of odd zeta values using Euler's reflection formula. The advantage of doing so will only be clear upon combination with results from the upper part. As we will only focus on computing the integral over $\mathcal{F}_L$ at five points, we refer to the complete four-point results, including the contributions from the upper part, to \cite{Claasen:2024ssh}. 

\subsection{The general five-point amplitude}
\label{ssec:gen5pt}
The minimal pure spinor formalism dictates that a type II five-point superstring amplitude can be written as \cite{Green:2013bza} 
\begin{align}
	\mathcal{M}^{\rm 5pt}_{\rm 1-loop}&=2\int_{\mathcal{F}}\frac{\text{d}^2\tau}{\tau_2^2}\Bigg(\prod_{k=2}^5\int_{\Sigma_\tau}\frac{d^2z_k}{\tau_2}\Bigg)\, \KN_5(s_{ij})\nonumber\\&\times\tau_2\Big\{\sum_{\substack{2\leq i<j \\ 2\leq k<l}}(f^{(1)}_{ij}\tilde{f}^{(1)}_{kl}-\frac{\pi}{\tau_2}\delta_{ik}\delta_{jl}s_{ij})\langle C_{1,ij}\tilde{C}_{1,kl}\rangle+\frac{\pi}{\tau_2}\langle C_{1,2,3,4,5}^m\tilde{C}_{1,2,3,4,5}^m\rangle\Big\}\,,
	\label{eq:5ptint}
\end{align}
with $\text{KN}_5(s_{ij})$ again the Koba-Nielsen factor (\ref{eq: integrandB}). The $f_{ij}^{(1)
}$ is shorthand for $f^{(1)}(z_i-z_j;\tau)$, which is related to the Green function (\ref{eq: greensfourier}) by $f^{(1)}(z;\tau)=-\partial_zG(z;\tau)$. The $C_{1,ij}$ are the open-string BRST invariants that can be expressed in terms of linear combinations of the two independent YM tree amplitudes bilinear in Mandelstams \cite{Mafra:2011nv,Mafra:2011nw}
\begin{align}
	\langle C_{1,23}\rangle=s_{45}\big(s_{24}A_{\text{YM}}(1,3,2,4,5)-s_{34}A_{\text{YM}}(1,2,3,4,5)\big)\,.
	\label{eq:opBRST}
\end{align}
This guarantees the presentability of $\langle C_{1,ij}\tilde{C}_{1,kl}\rangle$ in terms of the four independent kinematic factors $\{A_\ym(1,2,3,4,5),A_\ym(1,3,2,4,5)\}\times\{\tilde{A}_\ym(1,2,3,4,5),\tilde{A}_\ym(1,3,2,4,5)\}$. The other BRST invariant, $C_{1,2,3,4,5}^m\tilde{C}_{1,2,3,4,5}^m$ can also be matched with bilinear combinations of the $\ym$ tree amplitudes for any type IIB component expansion\footnote{In type IIA, this is no longer true due to the opposite space-time chirality of the left- and rightmovers. As detailed in \cite{Green:2013bza}, this causes the amplitude to be no longer expressible as bilinears in Yang-Mills trees.} \cite{Green:2013bza}. However, in this case, the expansion coefficients depend on the overall ${\rm U}(1)$-charge $q$ of the string states of the supergravity multiplet that are being scattered. One can show that the BRST invariant can be expressed as \cite{Mafra:2010pn,Green:2013bza}
\begin{align}
	\langle C_{1,2,3,4,5}^m\tilde{C}_{1,2,3,4,5}^m\rangle=\begin{cases}
		+1A_{\ym}^tS_0M_3\tilde{A}_\ym\,, \quad {\rm U}(1)\, \,\text{conserved,}\,\,q=0\,. \\
		-\tfrac{1}{3}A_{\ym}^tS_0M_3\tilde{A}_\ym\,, \quad {\rm U}(1)\, \,\text{max. violation,}\,\,q=\pm 2\,.
	\end{cases}
	\label{eq:BRST}
\end{align}
For an amplitude with total charge $q=\pm 1$ such as one $B$-field and four gravitons, this BRST invariant vanishes \cite{Green:2013bza}. Furthermore, the product $\langle C_{1,ij}\tilde{C}_{1,kl}\rangle$ of open-string BRST-invariant becomes anti-symmetric under the exchange of $C\leftrightarrow\tilde{C}$, which means that only imaginary cuspidal modular graph forms survive in the assembly of the amplitude. As integrals of cusp forms vanish over the fundamental domain of $\SL(2,\mathbb Z)$ \cite{rankin_1939,selberg1940bemerkungen}, amplitudes with $q=\pm 1$ are necessarily zero.

Substituting the BRST invariant kinematic factors (\ref{eq:opBRST}) and (\ref{eq:BRST}) into (\ref{eq:5ptint}), we see that for both the ${\rm U}(1)$-conserving and ${\rm U}(1)$-violating components of the genus-one five-point amplitude we find the structure
\begin{align}
	\mathcal{M}_{\text{1-loop}}^{{\rm 5pt},\,q}=A_\ym^t \mathcal{S}_{\text{1-loop}}^{q}\tilde{A}_\ym\,.
	\label{eq:1loopklt}
\end{align}
Due to the dependence of $\mathcal{S}_{\text{1-loop}}^{q}$ on the ${\rm U}(1)$-charge of this expression, we find that the one-loop amplitude admits KLT-like representations only after restricting to fixed ${\rm U}(1)$-charge sectors. We stress that this does not rule out the possibility of the existence of a universal one-loop KLT formula in terms of Yang-Mills tree amplitudes. In addition to the dependence on $q$, the $2\times 2$ matrix $\mathcal{S}_{\text{1-loop}}^{q}$ depends only on the Mandelstam variables $s_{ij}$ and thus captures the $\ap$-dependence. In the next sections, we expand and compute the low-energy expansion of the worldsheet integrals in $\mathcal{S}_{\text{1-loop}}^{q}$ in both the conserving and violating sector. Before we do so, we first comment on some general features of these low-energy expansions.\\
\indent The low-energy expansion of (\ref{eq:5ptint}) is realized by expanding the Koba-Nielsen factor for small momenta, just as we did in the four-point example\footnote{There is a subtlety in the low-energy expansion of (\ref{eq:5ptint}) whenever $i=k$ and $j=l$ in the sum. This is because the integrand has a pole in $s_{ij}$ for this configuration. To safely expand this contribution, we must first remove this pole whose residue in the example of $i=1,j=2$ is the four-point integral with momenta $k_1+k_2,k_3,k_4,k_5$. For more details, we refer to \cite{Green:2013bza}.\label{footnote: diag}}. Similarly to the four-point example, this expansion is only valid in a region of the moduli space where the associated genus-one Riemann surface does not degenerate. Hence, we use the same split of the fundamental domain by an auxiliary cut-off $L$ as in (\ref{eq:split}) and expand the integral (\ref{eq:5ptint}) only in the lower region $\mathcal{F}_L$ so that
\begin{align}
	\mathcal{S}_{\text{1-loop}}^{q}(s_{ij})=\mathcal{S}_{L}^{q}(L;s_{ij})+\mathcal{S}_{R}^{q}(L;s_{ij})\,.
	\label{eq: 1loopsplit}
\end{align}
Unfortunately, in contrast to the four-point case, there are no results in the literature on the integral in the region $\mathcal{F}_R$ for five-point processes, which we leave for future work. The rest of this work will therefore be concerned with the integral over $\mathcal{F}_L$.\\
\indent After expanding the Koba-Nielsen factor in (\ref{eq:5ptint}), the resulting $\tau$-dependent integrands can again be expressed in terms of modular graph forms. The graphs that appear have at most five vertices with possible multiple edges between any two vertices. These can be collected order by order in a systematic way and translated one by one to the iterated integrals introduced before using an unpublished extension of the \textsc{MGFtoBeqv} package written by the authors. The reason that necessitates an extension is the appearance of non-reducible 5-point topologies such as
\begin{equation}
	\label{eq: graph}
	\vcenter{\hbox{
			\begin{tikzpicture}[
				>=latex,
				state/.style={circle, inner sep=3pt, font=\large},
				edge/.style={->, thick, shorten <=3pt, shorten >=3pt},
				lbl/.style={fill=white, inner sep=1.5pt, font=\scriptsize}
				]
				
				\node[state] (n2) at (0, 0) {$2$};
				\node[state] (n5) at (2, 1) {$5$};
				\node[state] (n1) at (2, -1) {$1$};
				\node[state] (n4) at (4, 0) {$4$};
				\node[state] (n3) at (2, 2.8) {$3$};
				
				\draw[edge] (n2) to[bend left=35] node[lbl] {$(1,0)$} (n3);
				\draw[edge] (n2) to[bend right=6] node[lbl] {$(1,1)$} (n3);
				\draw[edge] (n3) to[bend left=20] node[lbl] {$(1,1)$} (n4);
				\draw[edge] (n1) to node[lbl] {$(1,1)$} (n2);
				\draw[edge] (n5) to node[lbl] {$(1,1)$} (n2);
				\draw[edge] (n1) to node[lbl] {$(1,1)$} (n4);
				\draw[edge] (n5) to node[lbl] {$(1,1)$} (n4);
				\draw[edge] (n1) to node[lbl] {$(0,1)$} (n5);
				
			\end{tikzpicture}
	}}\,,
\end{equation}
which had not been implemented before, as the calculation of the four-graviton amplitude only involved four-point topologies. Instead of introducing all new graph topologies and determining their properties and relations, we immediately go to the iterated integral language through the extended conversion algorithm. In this way, the $\ap$-expansion of $\mathcal{S}^q_{\text{1-loop}}$ will be given in terms of a particular product of kinematic matrices whose coefficients are integrals over $\mathcal{F}$ of linear combinations of iterated Eisenstein integrals.

\subsection{The U(1)-conserving sector}
\label{ssec: conserv}
First, we focus on the scattering processes in the ${\rm U}(1)$-conserving sector. We take the example of five gravitons, which is related to other processes in the same sector by supersymmetry. As detailed in the previous section, the calculation of the low-energy expansion is based on expanding the Koba-Nielsen factor in (\ref{eq:5ptint}) (where we keep in mind the treatment of the kinematic poles as explained in footnote \ref{footnote: diag}). Performing this expansion, we write everything in terms of modular graph forms and bilinears in Yang-Mills trees. Using the extended conversion algorithm \textsc{MGFtoBeqv}, we find the following decomposition of the $\mathcal{S}_L^{q}$ kernel in the conserving sector
\begin{align}
	\mathcal{S}_{L}^{q=0}=&S_0\Bigl(\Xi_3M_3+\Xi_5M_5+\Xi_{3,3}M_3^2+\Xi_7M_7+\Xi_{7^\prime}M_7^\prime+\Xi_{\{3,5\}}\{M_3,M_5\}+\Xi_{8^\prime}M_8^\prime+\Xi_9M_9 \nonumber\\&+\Xi_{3,3,3}M_3^3+\Xi_{9^\prime}M_9^\prime+\Xi_{9^{\prime\prime}}M_9^{\prime\prime}+\Xi_{\{3,7\}}\big((1-\beta)\{M_3,M_7\}+\beta\{M_5,M_5\})\big)\nonumber\\&+\big(\Xi_{10^\prime}+\beta\,\Xi_{\{3,7\}}\big)\big({\{M_3,M_7\}}-\{M_5,M_5\}\big)+\Xi_{\{3,7^\prime\}}\{M_3,M_7^\prime\}+\Xi_{10^{\prime\prime}}M_{10^{\prime\prime}}\nonumber\\&+\Xi_{10^{\prime\prime\prime}}M_{10^{\prime\prime\prime}}+\mathcal{O}(\alpha'^{11})\Bigr)\,,
	\label{eq:rescons}
\end{align}
which extends the results in \cite{Green:2013bza}. This expression demands a couple of remarks. First, the unprimed $M_{\bullet}$ matrices are the same kinematic matrices that appeared in the genus-zero five-point amplitude. These come together with the unprimed $\Xi_{\bullet}$ that appeared in the genus-one four-graviton amplitude (see section \ref{ssec:4ptrev}). Hence, these signify five-point interactions that factor into four-point interactions and thus have kinematic poles when combined with bilinears in $A_{\rm YM}$. Second, beyond these remnants of previously studied interactions, we see that the five-point amplitude requires additional kinematic matrices and coefficients from seventh order in $\ap$ onward. Such interactions are signaled by primed kinematic matrices $M^\prime_\bullet$ and their coefficients $\Xi_{\bullet'}$ respectively. Just like their unprimed counterparts, the entries of the primed matrix $M_k^\prime$ are degree-$k$ polynomials in Mandelstam invariants. Furthermore, the specific linear combination $\{M_3,M_7\}-\{M_5,M_5\}$ does not factor into four-point interactions, as it is free of poles when combined with the $A_{\rm YM}$. Hence, this combination is also understood to fall under the primed interactions. The appearance of such non-factorizable terms implies a class of $D^{2w-2}R^5$ interactions that do not occur at tree level. These matrices share many algebraic properties with their unprimed counterparts, like preserving BCJ and KK relations among the $A_\ym$ they act on \cite{BCJ}. Last, the one-parameter freedom in the four-point result (\ref{eq: s1loop}) emerges again in (\ref{eq:rescons}), but in a completely different way. Whereas at four points the two matrices $\{M_3,M_7\}$ and $\{M_5,M_5\}$ were equal, the same is not true at five points. Still, we have the freedom to change the kinematic matrix that comes with $\Xi_{\{3,7\}}$ due to the absence of poles in the combination $\{M_3,M_7\}-\{M_5,M_5\}$.

The explicit forms of the primed $M$-matrices can be found in the Mathematica file accompanying the arXiv submission. The $\Xi_\bullet$ coefficients and their primed versions again consist of integrals over $\mathcal{F}_L$ of linear combinations of iterated Eisenstein integrals and can be found in appendix \ref{appendix: beqvs} and in the Mathematica file. In order to evaluate those integrals, we can use the results from \cite{Doroudiani:2023bfw}. However, there is one complication compared to the four-point case. The basis of integrands introduced in \cite{Doroudiani:2023bfw} consists of modular functions that satisfy certain Laplace equations. The Laplace equation for one of these functions has eigenvalue zero. The integral of this particular function, denoted $F_{2,2,3}^{+(1)}$ in \cite{Dorigoni:2024oft}, cannot be handled using the methods in \cite{Doroudiani:2023bfw} and was left undetermined. At four points, this integral did not play a role. However, at five points, it appears in $\Xi_{\{3,7'\}},\Xi_{10'},\Xi_{10''}$ and $\Xi_{10'''}$. We managed to evaluate it numerically using a method detailed in the appendix \ref{sec:appendix}, but we were unable to uncover its nature using PSLQ where we tried a basis of $\{\frac{\zeta_7}{\zeta_{3}^2},\frac{\zeta'_n}{\zeta_n},\gamma_E,\log(\pi),\log(2),1\}$. We find
\begin{align}
	\omega=0.00016752391745287411575\dots\,.
	\label{eq: omega}
\end{align}
The results are 
\begin{align}
	&\Xi_{7^\prime}\approx-\frac{\pi^2\zeta_3^2}{9}\bigg(\log(4\pi L)-\frac{37}{12}+\bigg[\gamma_E-\frac{\zeta_{-3}^\prime}{\zeta_{-3}}-\frac{\zeta_3^{\prime}}{\zeta_3}\bigg]\bigg)\,,\nonumber\\
	&\Xi_{8^\prime}\approx\frac{7\pi^2\zeta_5}{150}\,,\nonumber\\
	&\Xi_{9^\prime}\approx\frac{\pi^2\zeta_3^2}{3}c_{9'}+\frac{64\pi^2\zeta_5}{567}\bigg(\log(4\pi L)-\frac{22061}{4608}+\bigg[\gamma_E-\frac{\zeta_5'}{\zeta_5}+\frac{79}{768}\frac{\zeta_{-5}'}{\zeta_{-5}}+\frac{847}{768}\frac{\zeta_{-1}'}{\zeta_{-1}}-\frac{847}{384}\frac{\zeta_{-3}'}{\zeta_{-3}}\bigg]\bigg)\,,\nonumber\\
	&\Xi_{9^{\prime\prime}}\approx\frac{\pi^2\zeta_3^2}{3}c_{9''}+\frac{31\pi^2\zeta_5}{2268}\bigg(\log(4\pi L)-\frac{6133}{1116}+\bigg[\gamma_E-\frac{\zeta_5'}{\zeta_5}-\frac{2}{93}\frac{\zeta_{-5}'}{\zeta_{-5}}+\frac{91}{93}\frac{\zeta_{-1}'}{\zeta_{-1}}-\frac{182}{93}\frac{\zeta_{-3}'}{\zeta_{-3}}\bigg]\bigg)\,,\nonumber\\
	&\Xi_{\{3,7'\}}\approx\frac{\pi^2\zeta_7}{3}c_{\{3,7'\}}-\frac{263221283\pi^2\zeta_3^2}{2277552060}\bigg(\log(4\pi L)+\frac{2251802584710}{263221283}\omega-\frac{41824531609}{1052885132000}\nonumber\\&\qquad+\bigg[-\frac{232456169}{263221283}\gamma_E+\frac{232456169}{263221283}\frac{\zeta_{3}'}{\zeta_{3}}+\frac{495677452}{263221283}\frac{\zeta_{-5}'}{\zeta_{-5}}-\frac{758898735}{263221283}\frac{\zeta_{-3}'}{\zeta_{-3}}\bigg]\bigg)\,,\nonumber\\
	&\Xi_{10^{\prime}}\approx\frac{\pi^2\zeta_7}{3}c_{10'}-\frac{451201987\pi^2\zeta_3^2}{5693880150}\bigg(\log(4\pi L)+\frac{793064385540}{451201987}\omega-\frac{258716762069}{94989892000}\nonumber\\&\qquad+\bigg[\frac{709219094}{451201987}\gamma_E-\frac{709219094}{451201987}\frac{\zeta_{3}'}{\zeta_{3}}-\frac{258017107}{451201987}\frac{\zeta_{-5}'}{\zeta_{-5}}-\frac{193184880}{451201987}\frac{\zeta_{-3}'}{\zeta_{-3}}\bigg]\bigg)\,,\nonumber\\
	&\Xi_{10^{\prime\prime}}\approx\frac{\pi^2\zeta_7}{3}c_{10''}+\frac{22321063\pi^2\zeta_3^2}{79714322100}\bigg(\log(4\pi L)-\frac{64032908940}{22321063}\omega-\frac{276650114659}{89284252000}\nonumber\\&\qquad+\bigg[\frac{32978936}{22321063}\gamma_E-\frac{32978936}{22321063}\frac{\zeta_{3}'}{\zeta_{3}}-\frac{10657873}{22321063}\frac{\zeta_{-5}'}{\zeta_{-5}}-\frac{11663190}{22321063}\frac{\zeta_{-3}'}{\zeta_{-3}}\bigg]\bigg)\,,\nonumber\\
	&\Xi_{10^{\prime\prime\prime}}\approx\frac{\pi^2\zeta_7}{3}c_{10'''}+\frac{44803\pi^2\zeta_3^2}{1184327071200}\bigg(\log(4\pi L)-\frac{54216540}{44803}\omega-\frac{450446939}{179212000}\nonumber\\&\qquad+\bigg[\frac{40046}{44803}\gamma_E-\frac{40046}{44803}\frac{\zeta_{3}'}{\zeta_{3}}+\frac{4757}{44803}\frac{\zeta_{-5}'}{\zeta_{-5}}-\frac{49560}{44803}\frac{\zeta_{-3}'}{\zeta_{-3}}\bigg]\bigg)\,,
	\label{eq: xipsint}
\end{align}
where $c_{\bullet'}\in\QQ$ are conjecturally rational numbers \cite{Zerbini:2015rss} that we were unable to determine due to lack of knowledge of the Laurent polynomials of MGFs of tetrahedral topology or total modular weight greater than 12. We have not rewritten the unprimed $\Xi_\bullet$ as they are the same as those in the four-point case \eqref{eq: xisint4pt}. One way in which these coefficients could be determined without calculating the Laurent polynomials of hundreds of MGFs would be to use the correspondence with UV-divergences of one-loop matrix elements in effective
field theory \cite{Edison:2021ebi}. Unfortunately, they do not provide data at the orders in $\ap$ where the $c_{\bullet'}$ are unknown. We leave the calculations of these orders for future work.

Note that $\Xi_{10''}$ and $\Xi_{10'''}$ were designed to capture all integrals of $\betaeqv{1\\4}\zeta_5$ and $\betaeqv{3\\8}\zeta_3$ respectively (see appendix \ref{appendix: beqvs}). This fixed the freedom in redefinition of the basis of $M$-matrices by simultaneously adapting the coefficients. The reason for this particular choice will be explained at the end of the following section.

\subsection{The U(1)-violating sector}
\label{ssec:viol}
Recall that the maximum amount of $\rm U(1)$ violation allowed for non-vanishing five-point amplitudes was found using the bound (\ref{eq: qmax}) to be $\lvert q\rvert_{\rm max}=2$. As amplitudes for $q=1$ vanish, we consider only the sector where $\lvert q\rvert=2$. Boels showed in \cite{Boels:2012zr,Boels:2013jua} that maximally R-symmetry violating (MRV) amplitudes in type IIB exhibit remarkable simplicity. In fact, the bilinears in $A_{\rm YM}$ for MRV components are all proportional to a single kinematic factor multiplying a totally symmetric function in Mandelstam variables. For the example of four gravitons together with one holomorphic axio-dilaton, we denote the kinematic factor by $K_5$. It is built from the $R^4$ matrix element and a rank sixteen tensor $t_{16}$ \cite{Oli:pc}
\begin{align}
	K_5\propto t_{16}^{m_1m_2\dots m_8n_1n_2\dots n_8}k_1^{m_1}k_1^{n_1}h_1^{m_2n_2}\dots k_4^{m_7}k_4^{n_7}h_4^{m_8n_8}\,,
	\label{eq: K5}
\end{align}
where $h_i^{mn}$ are the graviton polarizations. In the momentum phase space of four massless particles, the $t_{16}$ tensor factorizes into two copies of the well-known $t_8$ tensor. This is no longer true at five points. The bilinears in $A_{\rm YM}$ for MRV components then reduce to \cite{Boels:2013jua}
\begin{align}
	\tilde{A}_{\rm YM}(1,\rho,4,5)A_{\rm YM}(1,\sigma,4,5)=J_{\rho,\sigma}K_5\,,
	\label{eq: aymk5}
\end{align}
with
\begin{align}
	J=\frac{1}{s_{12}s_{13}s_{23}s_{24}s_{34}s_{45}s_{15}}\begin{pmatrix}
		s_{13}s_{24}&\frac{1}{2}(s_{14}s_{23}-s_{12}s_{34}-s_{13}s_{24})\\\frac{1}{2}(s_{14}s_{23}-s_{12}s_{34}-s_{13}s_{24})&s_{12}s_{34}
	\end{pmatrix} \, .
	\label{eq: jmatrix}
\end{align}
From these expressions it is not immediate that the final amplitude is expressible in terms of symmetric polynomials in the Mandelstams, but this can be explicitly shown in examples and is guaranteed by the Bose symmetry of the on-shell superfield \cite{Boels:2012zr}. A generating basis of the ring of such symmetric polynomials for the scattering of five particles was proposed in \cite{Boels:2013jua} and is given by 
\begin{align}
	\tau_k=s_{12}^k+\text{all permutations}\qquad k\in\{2,3,\dots,9\}\,,
	\label{eq: symmpoly}
\end{align}
together with an extra element at degree six
\begin{align}
	\tau_{6}'=s_{12}^3s_{23}^3+\text{all permutations}\,.
	\label{eq: addgen}
\end{align}
The amount of basis polynomials at each degree can be calculated using Molien's theorem (see \cite{stanley1979invariants,Boels:2013jua}). For convenience, we list them for four and five points up to degree 7 in table \ref{tab:molien}.
\begin{table}[]
	\centering
	\begin{tabular}{c|cccccccc}
		&0&1&2&3&4&5&6&7  \\
		\hline 4&1&0&1&1&1&1&2&1\\
		5&1&0&1&1&2&2&5&4
	\end{tabular}
	\caption{Numbers of completely symmetric polynomials for 4 and 5 particles, subject to momentum conservation for polynomial degrees 0 to 7.}
	\label{tab:molien}
\end{table}
%
%
Employing the basis introduced above, we find the following result
\begin{align}
	\tilde{A}_{\rm YM}& \mathcal{S}_{L}^{q=\pm2}A_{\rm YM}=K_5\bigg(-\Xi_3+\frac{1}{4}\tau_2\Xi_5-\frac{1}{3}\tau_3\Xi_{3,3}+\bigg[\frac{3}{64}\Xi_7-\frac{1}{12}\hat{\Xi}_{7'}\bigg]\tau_2^2+\bigg[\frac{9}{16}\Xi_7-\frac{1}{3}\hat{\Xi}_{7'}\bigg]\tau_4\nonumber\\&+\bigg[-\frac{5}{24}\Xi_{\{3,5\}}+\frac{5}{36}\hat{\Xi}_{8'}\bigg]\tau_2\tau_3+\bigg[-\frac{3}{10}\Xi_{\{3,5\}}-\frac{1}{3}\hat{\Xi}_{8'}\bigg]\tau_5+\bigg[\frac{3965}{7776}\Xi_{9}-\frac{25}{72}\Xi_{3,3,3}+\frac{1}{6}\hat{\Xi}_{9'}\nonumber\\&-\hat{\Xi}_{9''}-\frac{5}{12}\hat{\Xi}_{9'''}\bigg]\tau_2\tau_4+\bigg[-\frac{35}{486}\Xi_{9}+\frac{5}{18}\Xi_{3,3,3}-\frac{1}{9}\hat{\Xi}_{9'}+\frac{4}{9}\hat{\Xi}_{9''}+\frac{2}{9}\hat{\Xi}_{9'''}\bigg]\tau_3^2+\bigg[\frac{65}{486}\Xi_{9}\nonumber\\&+\frac{5}{18}\Xi_{3,3,3}-\frac{2}{9}\hat{\Xi}_{9'}+\frac{8}{9}\hat{\Xi}_{9''}+\frac{2}{9}\hat{\Xi}_{9'''}\bigg]\tau_6+\bigg[-\frac{1235}{31104}\Xi_{9}+\frac{5}{144}\Xi_{3,3,3}+\frac{1}{12}\hat{\Xi}_{9''}\nonumber\\&+\frac{1}{24}\hat{\Xi}_{9'''}\bigg]\tau_2^3+\bigg[\frac{70}{243}\Xi_{9}-\frac{5}{9}\Xi_{3,3,3}+\frac{4}{9}\hat{\Xi}_{9'}-\frac{16}{9}\hat{\Xi}_{9''}-\frac{7}{9}\hat{\Xi}_{9'''}\bigg]\tau_6'+\bigg[\frac{1}{2}\hat{\Xi}_{10'}\nonumber\\&-349336\hat{\Xi}_{10'''}\bigg]\tau_2\tau_5+\bigg[-\frac{21}{80}\Xi_{\{3,7\}}+\frac{1}{9}\hat{\Xi}_{\{3,7'\}}-\frac{7}{16}\hat{\Xi}_{10'}+\frac{5553977}{9}\hat{\Xi}_{10'''}\bigg]\tau_3\tau_4\nonumber\\&+\bigg[-\frac{9}{35}\Xi_{\{3,7\}}-\frac{3}{7}\hat{\Xi}_{10'}+\frac{83408}{7}\hat{\Xi}_{10'''}\bigg]\tau_7+\bigg[-\frac{7}{320}\Xi_{\{3,7\}}-\frac{1}{36}\hat{\Xi}_{\{3,7'\}}-\frac{7}{192}\hat{\Xi}_{10'}\nonumber\\&+\frac{376493}{36}\hat{\Xi}_{10'''}\bigg]\tau_2^2\tau_3+\mathcal{O}\ap^{11})\bigg)\,.
	\label{eq:violboels}
\end{align}
It is clear from this expression that at each order in $\ap$ (i.e. polynomial degree in $s_{ij}$), the number of basis elements coincides with table \ref{tab:molien}. The new $\hat{\Xi}_{\bullet'}$ are given by
\begin{align}
	&\hat{\Xi}_{7'}\approx-\frac{\pi^2\zeta_3}{5}\bigg(\log(4\pi L)-\frac{127}{36}+\bigg[\gamma_E-\frac{\zeta_{-3}^\prime}{\zeta_{-3}}-\frac{\zeta_3^{\prime}}{\zeta_3}\bigg]\bigg)\,,\nonumber\\
	&\hat{\Xi}_{8'}\approx -\frac{101\pi^2 \zeta_5}{100}\,,\nonumber\\
	&\hat{\Xi}_{9'}\approx \hat{c}_{9'}\frac{\pi^2\zeta_3^2}{3}-\frac{304\pi^2\zeta_5}{5103}\bigg(\log(4\pi L)+\frac{919517}{109440}\nonumber\\&\qquad+\bigg[\gamma_E-\frac{\zeta_{5}'}{\zeta_5}-\frac{17459}{3648}\frac{\zeta'_{-5}}{\zeta_{-5}}-\frac{13811}{3648}\frac{\zeta_{-1}'}{\zeta_{-1}}+\frac{13811}{1824}\frac{\zeta_{-3}'}{\zeta_{-3}}\bigg]\bigg)\,,\nonumber\\
	&\hat{\Xi}_{9''}\approx \hat{c}_{9''}\frac{\pi^2\zeta_3^2}{3}-\frac{4645\pi^2\zeta_5}{20412}\bigg(\log(4\pi L)-\frac{1873979}{836100}\nonumber\\&\qquad+\bigg[\gamma_E-\frac{\zeta_{5}'}{\zeta_5}+\frac{2494}{13935}\frac{\zeta'_{-5}}{\zeta_{-5}}+\frac{16429}{13935}\frac{\zeta_{-1}'}{\zeta_{-1}}-\frac{32858}{13935}\frac{\zeta_{-3}'}{\zeta_{-3}}\bigg]\bigg)\,,\nonumber\\
	&\hat{\Xi}_{9'''}\approx  \hat{c}_{9'''}\frac{\pi^2\zeta_3^2}{3}+\frac{88\pi^2\zeta_5}{189}\bigg(\log(4\pi L)-\frac{2861}{990}+\bigg[\gamma_E-\frac{\zeta_{5}'}{\zeta_5}+\frac{2}{33}\frac{\zeta'_{-5}}{\zeta_{-5}}+\frac{35}{33}\frac{\zeta_{-1}'}{\zeta_{-1}}-\frac{70}{33}\frac{\zeta_{-3}'}{\zeta_{-3}}\bigg]\bigg)\,,\nonumber\\
	&\hat{\Xi}_{\{3,7'\}}\approx\hat{c}_{\{3,7'\}}\frac{\pi^2\zeta_7}{3}-\frac{90438073\pi^2\zeta_3^2}{76198500}\bigg(\log(4\pi L)-\frac{50342703390}{90438073}\omega -\frac{2861750531767}{1085256876000}\nonumber\\&\qquad+\bigg[\frac{123513001}{90438073}\gamma_E -\frac{123513001}{90438073}\frac{\zeta_3'}{\zeta_3}-\frac{33074928}{90438073}\frac{\zeta_{-5}'}{\zeta_{-5}}-\frac{57363145}{90438073}\frac{\zeta_{-3}'}{\zeta_{-3}}\bigg]\bigg)\,,\nonumber\\
	&\hat{\Xi}_{10'}\approx\hat{c}_{10'}\frac{\pi^2\zeta_7}{3}-\frac{1190393\pi^2\zeta_3^2}{5442750}\bigg(\log(4\pi L)-\frac{1665592740}{1190393}\omega-\frac{240046687429}{99993012000}\nonumber\\&\qquad+\bigg[\frac{1553066}{1190393}\gamma_E -\frac{1553066}{1190393}\frac{\zeta_3'}{\zeta_3}-\frac{362673}{1190393}\frac{\zeta_{-5}'}{\zeta_{-5}}-\frac{827720}{1190393}\frac{\zeta_{-3}'}{\zeta_{-3}}\bigg]\bigg)\,,\nonumber\\
	&\hat{\Xi}_{10'''}\approx \hat{c}_{10''}\frac{\pi^2\zeta_7}{3}+\frac{73\pi^2\zeta_3^2}{37736400}\bigg(\log(4\pi L)+\frac{49140}{73}\omega-\frac{45253007}{18396000}\nonumber\\&\qquad+\bigg[\frac{278}{219}\gamma_E-\frac{278}{219}\frac{\zeta_3'}{\zeta_3}-\frac{59}{219}\frac{\zeta_{-5}'}{\zeta_{-5}}-\frac{160}{219}\frac{\zeta_{-3}'}{\zeta_{-3}}\bigg]\bigg)\,.
	\label{eq: xihpsint}
\end{align}
\indent The reason for the combinations of $\Xi$-coefficients in (\ref{eq:violboels}) is the following. As an alternative basis, we can take the kinematic matrices $M_\bullet$ and $M_{\bullet }'$ from the conserving sector and multiply them with the bilinear in $A_{\rm YM}$ and the field theory limit $S_0$ to ``project'' them down to symmetric polynomials in the Mandelstams. This is the notion of projection we alluded to at the end of section \ref{ssec: conserv}. We denote this projection with by a linear operator $\mathcal{P}(M_{\bullet})=\tilde{A}_{\rm YM}S_0 M_\bullet A_{\rm YM}$, where $A_{\rm YM}$ and $\tilde{A}_{\rm YM}$ are in their $\mathrm{U}(1)$-violating state configurations. This allows us to write down the result in the MRV sector as 
\begin{align}
	\tilde{A}&_{\rm YM} \mathcal{S}_{L}^{q=\pm2}A_{\rm YM}\!=\!\mathcal{P}\Bigl(-\frac{1}{3}\Xi_3{M}_3+\frac{1}{5}\Xi_5{M}_5+\frac{1}{3}\Xi_{3,3}{M}_3^2+\frac{3}{7}\Xi_7{M}_7+\hat{\Xi}_{7^\prime}{M}_7^\prime+\frac{1}{2}\Xi_{\{3,5\}}\{M_3,M_5\}\nonumber\\&+\hat{\Xi}_{8^\prime}M_8^\prime+\frac{5}{9}\Xi_9M_9 +\frac{5}{9}\Xi_{3,3,3}M_3^3+\hat{\Xi}_{9^\prime}M_9^\prime+\hat{\Xi}_{9^{\prime\prime}}M_9^{\prime\prime}+\hat{\Xi}_{9^{\prime\prime\prime}}M_9^{\prime\prime\prime}+ \hat{\Xi}_{\{3,7^\prime\}}\{M_3,M_7^\prime\}\nonumber\\&+\frac{3}{5}\Xi_{\{3,7\}}\big((1-\beta)\{M_3,M_7\}+\beta\{M_5,M_5\})\big)+\big(\hat{\Xi}_{10^\prime}+\frac{3\beta}{5}\Xi_{\{3,7\}}\big)\big({\{M_3,M_7\}}-\{M_5,M_5\}\big)\nonumber\\&+\hat{\Xi}_{10^{\prime\prime\prime}}M_{10}^{\prime\prime\prime}+\mathcal{O}(\alpha'^{11})\Bigr)\,,
	\label{eq:resviol}
\end{align}
where $M_9'''$ has no counterpart in the conserving sector and should not be understood as another kinematic matrix. Rather, it is an extra symmetric polynomial required to describe the MRV sector completely and is only defined as such
\begin{align}
	\mathcal{P}(M_{9}''')=-\frac{5}{12}\tau_2\tau_4+\frac{2}{9}\tau_3^2+\frac{2}{9}\tau_6+\frac{1}{24}\tau_2^3-\frac{7}{9}\tau_6'\,.
\end{align}
The projection of all $M$-matrices can be found in appendix \ref{appendix: symPol}. Another point worth noting is that describing the conserving sector at tenth order in $\ap$ required a total of five kinematic structures. According to table \ref{tab:molien}, the basis of symmetric polynomials of degree 7 relevant to the MRV sector is only four-dimensional. This means that if we use the $M_\bullet$ and $M_\bullet '$ from the conserving sector as a starting point and project, we can only get four independent symmetric polynomials. The matrices in the conserving sector were engineered so that the projection of $M_{10}''$ does not appear in \eqref{eq:resviol}. 

\subsection{Connecting sectors through $S$-duality}
\label{ssec: sdual}
In section \ref{ssec: sRtypeIIB}, we showed how contributions to low-energy expansions of perturbative type IIB superstring amplitudes at four points fit into $S$-duality invariant coefficient functions $\mathcal{E}_\bullet$. In particular, these terms contribute to the zero mode of $\mathcal{E}_\bullet$. In this section, we investigate how contributions from five-point scattering processes fit into this picture.

When writing down the $N$-point effective action, we need notation to distinguish invariants contained in interactions such as $D^{2w}R^4$ from those in $D^{2w-2}R^5$ whenever $w>3$. We use the same notation as in \cite{Green:2013bza}
\begin{itemize}
	\item The family of interactions giving rise to products of $M$-matrices $M_{p_1}M_{p_2}\dots M_{p_n^\prime}M_{p_{n+1}^\prime}\dots$ is denoted by $(\oplus_{\ell=0}^w D^{2w-2\ell}R^{4+\ell})_{p_1,p_2,\dots,p_n^\prime,p_{n+1}^\prime,\dots}$
	\item For (anti-)commutators of $M$-matrices, we use the notation $(\oplus_{\ell=0}^w D^{2w-2\ell}R^{4+\ell})_{[p_1,p_2]\dots}$ and $(\oplus_{\ell=0}^w D^{2w-2\ell}R^{4+\ell})_{\{p_1,p_2\}\dots}$ and similarly for (anti-)commutators of primed $M$-matrices.
\end{itemize}

The perturbative results in the conserving sector for four and five points (\ref{eq: xisint4pt}) and (\ref{eq: xipsint}) imply that the four-point coefficient functions $\mathcal{E}_{\bullet}$ from section \ref{ssec: sRtypeIIB} come with $(D^{2w}R^4\oplus D^{2w-2}R^5)_{p_1,\dots}$. This is because we retrieve the same kinematic structures at five points up to $w=7$. To extract the interactions with powers of $R$ higher than 5, one must analyze the low-energy expansion of higher-point amplitudes. For example, $D^{2w-4} R^6$ and $D^{2w-6} R^7$ terms correspond to six and seven points (e.g. see \cite{Mafra:2016nwr} for the six-point calculations). It would be interesting to see how these higher-point interactions fit into this observation.

Beyond the $\mathcal{E}_\bullet$ remnants from four points that come with five-point tree-level kinematic structures, we find new kinematic structures that are absent in four-point tree-level computations. This signals the existence of new $S$-duality invariant coefficients $\mathcal{E}_{\bullet'}$. Consequently, these functions must contain perturbative terms that start at one loop and are associated to local five-field operators. Up to tenth order, we summarize all data in table \ref{tab:consint}.
\begin{table}[htbp]
	\renewcommand{\arraystretch}{1.13}
	\setlength{\tabcolsep}{5pt}
	\centering
	\begin{tabular}{|c|c|c|c|c|}
		\hline $\boldsymbol{M_\bullet}$ & \textbf{Interactions} & \textbf{Coeff.} & \textbf{Tree} & \textbf{1-loop} \\
		\hline 1 & $R$ & 1 & 1 & 0 \\
		\hline $M_3$ & $(R^4)_3$ & $\mathcal{E}_3$ & $2\zeta_3$ & $\Xi_3$ \\
		\hline $M_5$ & $(D^4R^4+D^2R^5)_5$ & $\mathcal{E}_5$ & $2\zeta_5$ & $\Xi_5$ \\
		\hline $M_3^2$ & $(D^6R^4+D^4R^5)_{3,3}$ & $\mathcal{E}_{3,3}$ & $2\zeta_3^2$ & $\Xi_{3,3}$ \\
		\hline $M_7$ & $(D^8R^4+D^6R^5)_7$ & $\mathcal{E}_7$ & $2\zeta_7$ & $\Xi_7$ \\
		\hline $M_{7}'$ & $(D^6R^5)_{7'}$ & $\mathcal{E}_{7'}$ & $0$ & $\Xi_{7'}$ \\
		\hline $\{M_3,M_5\}$ & $(D^{10}R^4+D^8R^5)_{\{3,5\}}$ & $\mathcal{E}_{\{3,5\}}$ & $2\zeta_3\zeta_5$ & $\Xi_{\{3,5\}}$ \\
		\hline $M_8'$ & $(D^8R^5)_{8'}$ & $\mathcal{E}_{8'}$ & $0$ & $\Xi_{8'}$ \\
		\hline $M_9$ & $(D^{12}R^4+D^{10}R^5)_{9}$ & $\mathcal{E}_9$ & $2\zeta_9$ & $\Xi_9$ \\
		\hline $M_{3}^3$ & $(D^{12}R^4+D^{10}R^5)_{3,3,3}$ & $\mathcal{E}_{3,3,3}$ & $\frac{4}{3}\zeta_3^3$ & $\Xi_{3,3,3}$ \\
		\hline $M_9'$ & $(D^{10}R^5)_{9'}$ & $\mathcal{E}_{9'}$ & $0$ & $\Xi_{9'}$ \\
		\hline $M_9''$ & $(D^{10}R^5)_{9''}$ & $\mathcal{E}_{9''}$ & $0$ & $\Xi_{9''}$ \\
		\hline \begin{tabular}{@{}c@{}}$(1\!-\!\beta)\{M_3,M_7\}$ \\ $+\beta\{M_5,M_5\}$\end{tabular} & \begin{tabular}{@{}c@{}}$(D^{14}R^4+D^{12}R^5)_w$ \\ \small{$w={(1\!-\!\beta)\{3,7\}\!+\!\beta\{5,5\}}$}\end{tabular} & $\mathcal{E}_{\{3,7\}}$ & \begin{tabular}{@{}c@{}}$2\zeta_3\zeta_7+\zeta_5^2$\end{tabular} & $\Xi_{\{3,7\}}$ \\
		\hline \begin{tabular}{@{}c@{}}$\{M_3,M_7\}$ \\ $-\{M_5,M_5\}$\end{tabular} & $(D^{12}R^5)_{10'}$ & $\mathcal{E}_{10'}$ & $2\beta\zeta_3\zeta_7-(1-\beta)\zeta_5^2$ & $\Xi_{10'}+\beta\Xi_{\{3,7\}}$ \\
		\hline $\{M_3,M_7'\}$ & $(D^{12}R^5)_{\{3,7'\}}$ & $\mathcal{E}_{\{3,7'\}}$ & $0$ & $\Xi_{\{3,7'\}}$ \\
		\hline $M_{10}''$ & $(D^{12}R^5)_{10''}$ & $\mathcal{E}_{10''}$ & $0$ & $\Xi_{10''}$ \\
		\hline $M_{10}'''$ & $(D^{12}R^5)_{10'''}$ & $\mathcal{E}_{10'''}$ & $0$ & $\Xi_{10'''}$ \\   
		\hline
	\end{tabular}
	\caption{Summary of tree-level and one-loop contributions to the ten-dimensional low-energy effective action arising from five-point interactions in the ${\rm U}(1)$-conserving sector.}
	\label{tab:consint}
\end{table}
The first column represents the kinematic structures associated to the interactions in the second column. The third column contains the $S$-duality invariant coefficients functions of the moduli, whose partial zero mode contributions arising from tree-level and one-loop computations can be found in column four and five respectively. This data fits into the schematic effective action as
\begin{align}
	S_{\text{eff}}^{q=0}\bigg\rvert_{\rm local}&=\int d^{10}x\sqrt{-G}\bigg(R+\mathcal{E}_3(R^4)_3+\mathcal{E}_5(D^4R^4+D^2R^5)_5+\mathcal{E}_{3,3}(D^6R^4+D^4R^5)_{3,3}\nonumber\\&+\mathcal{E}_{7}(D^8R^4+D^6R^5)_7+\mathcal{E}_{7^{\prime}}(D^6R^5)_{7^{\prime}}+\mathcal{E}_{\{3,5\}}(D^{10}R^4+D^8R^5)_{\{3,5\}}+\mathcal{E}_{8^{\prime}}(D^8R^5)_{8^{\prime}}\nonumber\\&+\mathcal{E}_{9}(D^{12}R^4+D^{10}R^5)_9+\mathcal{E}_{3,3,3}(D^{12}R^4+D^{10}R^5)_{3,3,3}+\mathcal{E}_{9^{\prime}}(D^{10}R^5)_{9^{\prime}}\nonumber\\&+\mathcal{E}_{9^{\prime\prime}}(D^{10}R^5)_{9^{\prime\prime}}+\mathcal{E}_{\{3,7 \}}(D^{14}R^4+D^{12}R^5)_{(1-\beta){\{3,7\}}+\beta\{5,5\}}+\mathcal{E}_{10'}(D^{12}R^5)_{\{3,7\}-\{5,5\}}\nonumber\\&+\mathcal{E}_{\{3,7'\}}(D^{12}R^5)_{\{3,7'\}}+\mathcal{E}_{10''}(D^{12}R^5)_{10''}+\mathcal{E}_{10'''}(D^{12}R^5)_{10'''}+\mathcal{O}(\ap^{11})\bigg)\,,
	\label{eq: effactcons}
\end{align}
where $G$ is the determinant of the space-time metric. Although we do not know the modular invariant coefficient functions $\mathcal{E}_\bullet$ and $\mathcal{E}_{\bullet '}$ beyond the ones identified in section \ref{ssec: sRtypeIIB}, we expect them to arise and we know that they must contain the perturbative contributions summarized in table \ref{tab:consint}. The fact that the relative coefficients between $D^{2w}R^4$ and $D^{2w-2}R^5$ are always one matches that they are part of the same supermultiplet and hence they should be associated with the same modular functions.

The ${\rm U}(1)$-violating sector can be analyzed in a similar manner. Recall that for five-point amplitudes $\lvert q\rvert_{\rm max}=2$, which means that the local operators transform as non-holomorphic modular forms of weights $(\mp1,\pm1)$. In order to ensure $S$-duality invariance, the associated coefficient functions must transform as non-holomorphic modular forms in the opposite manner, i.e. with weights $(\pm 1,\mp 1)$. In section \ref{ssec: sRtypeIIB}, we saw that contributions to the zero modes of the coefficient of $\phi R^4$ could be related to the ones of $\mathcal{E}_3$ associated to $R^4$ by acting with a covariant derivative operator. It is tempting to speculate that this pattern would continue for higher-order corrections. Concretely, this would mean the following. Let the zero mode of the Fourier expansion of $\mathcal{E}_\bullet$ of a ${\rm U}(1)$-conserving interaction $D^{2w}R^4$ be denoted 
\begin{align}
	a^{(w)}_0(\Omega_2)=:a_w\Omega_2^{(w+3)/2}+b_w\Omega_2^{(w-1)/2}+\mathcal{O}\big(\Omega^{(w-5)/2}\big)\,.
	\label{eq: zm}
\end{align}
Then, we act with the Maass raising operator to get a form $\nabla \mathcal{E}_\bullet$ of weights $(1,-1)$, whose zero mode is
\begin{align}
	\nabla a_0^{(w)}&=a_w\frac{w+3}{2}\Omega_2^{(w+3)/2}+b_w\frac{w-1}{2}\Omega_2^{(w-1)/2}+\mathcal{O}\big(\Omega^{(w-5)/2}\big)\nonumber\\&=:A_w\Omega_2^{(w+3)/2}+B_w\Omega_2^{(w-1)/2}+\mathcal{O}\big(\Omega^{(w-5)/2}\big)\,.
	\label{eq: nabzm}
\end{align}
The ratio of tree-level and one-loop amplitudes for $q=-2$ is related to that of $q=0$ by
\begin{align}
	\frac{B_w/A_w}{b_w/a_w}=\frac{w-1}{w+3}\,.
	\label{eq: ratio}
\end{align}
This pattern is exactly what we observe in the coefficients of $\Xi_\bullet$ in (\ref{eq:resviol}) and was already observed in \cite{Green:2013bza}. As a main new result, our analysis shows that this pattern continues to hold to the tenth order. We summarize again all the data in table \ref{tab:violint}.
\begin{table}[htbp]
	\renewcommand{\arraystretch}{1.13}
	\setlength{\tabcolsep}{5pt}
	\centering
	\begin{tabular}{|c|c|c|c|c|}
		\hline $\boldsymbol{\mathcal{P}(M_\bullet)}$ & \textbf{Interactions} & \textbf{Coeff.} & \textbf{Tree} & \textbf{1-loop} \\
		\hline 1 & $\times$ & 0 & 0 & 0 \\
		\hline $\mathcal{P}(M_3)$ & $(\phi R^4)_3$ & $\nabla^{(0)} \mathcal{E}_3$ & $2\zeta_3$ & $-\frac{1}{3}\Xi_3$ \\
		\hline $\mathcal{P}(M_5)$ & $(D^4\phi R^4)_5$ & $\nabla^{(0)}\mathcal{E}_5$ & $2\zeta_5$ & $\Xi_5$ \\
		\hline $\mathcal{P}(M_3^2)$ & $(D^6\phi R^4)_{3,3}$ & $\nabla^{(0)}\mathcal{E}_{3,3}$ & $2\zeta_3^2$ & $\frac{1}{3}\Xi_{3,3}$ \\
		\hline $\mathcal{P}(M_7)$ & $(D^8\phi R^4)_7$ & $\nabla^{(0)}\mathcal{E}_7$ & $2\zeta_7$ & $\frac{3}{7}\Xi_7$ \\
		\hline $\mathcal{P}(M_{7}')$ & $(D^8 \phi R^4)_{7'}$ & $\hat{\mathcal{E}}_{7'}$ & $0$ & $\hat{\Xi}_{7'}$ \\
		\hline $\mathcal{P}(\{M_3,M_5\})$ & $(D^{10}\phi R^4)_{\{3,5\}}$ & $\nabla^{(0)}\mathcal{E}_{\{3,5\}}$ & $2\zeta_3\zeta_5$ & $\frac{1}{2}\Xi_{\{3,5\}}$ \\
		\hline $\mathcal{P}(M_8')$ & $(D^{10} \phi R^4)_{8'}$ & $\hat{\mathcal{E}}_{8'}$ & $0$ & $\hat{\Xi}_{8'}$ \\
		\hline $\mathcal{P}(M_9)$ & $(D^{12}\phi R^4)_{9}$ & $\nabla^{(0)}\mathcal{E}_9$ & $2\zeta_9$ & $\frac{5}{9}\Xi_9$ \\
		\hline $\mathcal{P}(M_{3}^3)$ & $(D^{12}\phi R^4)_{3,3,3}$ & $\nabla^{(0)}\mathcal{E}_{3,3,3}$ & $\frac{4}{3}\zeta_3^3$ & $\frac{5}{9}\Xi_{3,3,3}$ \\
		\hline $\mathcal{P}(M_9')$ & $(D^{12}\phi R^4)_{9'}$ & $\hat{\mathcal{E}}_{9'}$ & $0$ & $\hat{\Xi}_{9'}$ \\
		\hline $\mathcal{P}(M_9'')$ & $(D^{12}\phi R^4)_{9''}$ & $\hat{\mathcal{E}}_{9''}$ & $0$ & $\hat{\Xi}_{9''}$ \\
		\hline $\mathcal{P}(M_{9}''')$ & $(D^{12} \phi R^4)_{9'''}$ & $\hat{\mathcal{E}}_{9'''}$ & $0$ & $\hat{\Xi}_{9'''}$ \\
		\hline \begin{tabular}{@{}c@{}}$\mathcal{P}\big((1\!-\!\beta)\{M_3,M_7\}$ \\ $+\beta\{M_5,M_5\}\big)$\end{tabular} & \begin{tabular}{@{}c@{}}$(D^{14} \phi R^4)_w$ \\ \small{$w={(1\!-\!\beta)\{3,7\}\!+\!\beta\{5,5\}}$}\end{tabular} & $\nabla^{(0)}\mathcal{E}_{\{3,7\}}$ & \begin{tabular}{@{}c@{}}$2\zeta_3\zeta_7+\zeta_5^2$\end{tabular} & $\frac{3}{5}\Xi_{\{3,7\}}$ \\
		\hline \begin{tabular}{@{}c@{}}$\mathcal{P}(\{M_3,M_7\}$ \\ $-\{M_5,M_5\})$\end{tabular} & $(D^{14} \phi R^4)_{10'}$ & $\hat{\mathcal{E}}_{10'}$ & $2\beta\zeta_3\zeta_7-(1-\beta)\zeta_5^2$ & \begin{tabular}{@{}c@{}}$\hat{\Xi}_{10'}$ \\ $+\frac{3\beta}{5}\Xi_{\{3,7\}}$\end{tabular} \\
		\hline $\mathcal{P}(\{M_3,M_7'\})$ & $(D^{14} \phi R^4)_{\{3,7'\}}$ & $\hat{\mathcal{E}}_{\{3,7'\}}$ & $0$ & $\hat{\Xi}_{\{3,7'\}}$ \\
		\hline $\mathcal{P}(M_{10}'')$ & $\times$ & $0$ & $0$ & $0$ \\
		\hline $\mathcal{P}(M_{10}''')$ & $(D^{14} \phi R^4)_{10'''}$ & $\hat{\mathcal{E}}_{10'''}$ & $0$ & $\hat{\Xi}_{10'''}$ \\   
		\hline
	\end{tabular}
	\caption{Summary of tree-level and one-loop contributions to the ten-dimensional low-energy effective action arising from five-point interactions in the ${\rm U}(1)$-violating sector.}
	\label{tab:violint}
\end{table}
It fits into the effective action
\begin{align}
	S_{\rm eff}^{q=2}\bigg\rvert_{\rm local}&=\int d^{10}x\sqrt{-G}\bigg(\nabla^{(0)}\mathcal{E}_3(\phi R^4)_3+\nabla^{(0)}\mathcal{E}_5(D^4\phi R^4)_5+\nabla^{(0)}\mathcal{E}_{3,3}(D^6 \phi R^4)_{3,3}\nonumber\\&+\nabla^{(0)}\mathcal{E}_{7}(D^8 \phi R^4)_7+\hat{\mathcal{E}}_{7^{\prime}}(D^8 \phi R^4)_{7^{\prime}}+\nabla^{(0)}\mathcal{E}_{\{3,5\}}(D^{10}\phi R^4)_{\{3,5\}}+\hat{\mathcal{E}}_{8^{\prime}}(D^{10}\phi R^4)_{8^{\prime}}\nonumber\\&+\nabla^{(0)}\mathcal{E}_{9}(D^{12}\phi R^4)_9+\nabla^{(0)}\mathcal{E}_{3,3,3}(D^{12}\phi R^4)_{3,3,3}+\hat{\mathcal{E}}_{9^{\prime}}(D^{12}\phi R^4)_{9^{\prime}}+\hat{\mathcal{E}}_{9''}(D^{12}\phi R^4)_{9''}\nonumber\\&+\hat{\mathcal{E}}_{9'''}(D^{12}\phi R^4)_{9'''}+\nabla^{(0)}\mathcal{E}_{\{3,7 \}}(D^{14}\phi R^4)_{(1-\beta){\{3,7\}}+\beta\{5,5\}}+\hat{\mathcal{E}}_{10'}(D^{14}\phi R^4)_{\{3,7\}-\{5,5\}}\nonumber\\&+\hat{\mathcal{E}}_{\{3,7'\}}(D^{14}\phi R^4)_{\{3,7'\}}+\hat{\mathcal{E}}_{10'''}(D^{14}\phi R^4)_{10'''}+\mathcal{O}(\ap^{11})\bigg)\,.
	\label{eq: effactviol}
\end{align}
As explained in the previous section, $\mathcal{P}(M_{10}'')$ was engineered such that it does not appear here and to conform to the basis counting based on table \ref{tab:molien}. This explains the absence of the associated interaction in table \ref{tab:violint} and equation (\ref{eq: effactviol}).

\subsection{Comments on transcendentality patterns}
\label{ssec: comments}
Even though all results obtained are only the contributions of the integrations over the lower part $\mathcal{F}_L$ of the moduli space of the torus, there are still several recurring patterns in equations (\ref{eq: xisint4pt}), (\ref{eq: xipsint}), and (\ref{eq: xihpsint}). First of all, the terms with highest transcendental weight always arise from the constant terms of the zero modes of both the modular graph forms and the equivariant iterated Eisenstein integrals. The transcendental weight of these terms agrees with the order of $\ap$ minus three and therefore conforms uniform transcendentality. The terms involving zeta values of subleading transcendental weight arise from the integrals of the equivariant iterated Eisenstein integrals over the fundamental domain. Strikingly, if we pull out the overall prefactor of the $\log(4\pi L)$ term, we always find within the square brackets of (\ref{eq: xisint4pt}), (\ref{eq: xipsint}), and (\ref{eq: xihpsint}) affine linear combination of the following form 
\begin{align}
	\sum_{n\in N}\ c_{n}f_n\,,\qquad f_n\in\bigg\{-\frac{\zeta_n'}{\zeta_n},-\gamma_E\bigg\}\,,\qquad \sum_{n\in N}c_n=1\,,
	\label{eq: affine}
\end{align}
where $N$ is a finite set of odd integers excluding 1. Interestingly, if we include the unknown number $\omega$ in the square brackets, this pattern breaks, suggesting that we should treat $\omega$ on different footing compared to $\gamma_E$ and $\frac{\zeta_n'}{\zeta_n}$. We want to stress that this property holds for all 18 expressions we found and calls for an explanation. Unfortunately, the authors were unable to identify the reason for this pattern. We suggest that it can be used as a sanity check of any future, e.g., higher-point calculations. A possible reason why $\gamma_E$ should be included in the set in (\ref{eq: affine}) and treated on the same footing as $\frac{\zeta_n'}{\zeta_n}$ is because we can write $\gamma_E$ as the regular part of $\frac{\zeta_1'}{\zeta_1}$. That is, we have 
\begin{align}
	\gamma_E=\lim_{n\rightarrow 1}\bigg(\frac{\zeta_n'}{\zeta_n}+\frac{1}{n-1}\bigg)\,.
	\label{eq: eulergamma}
\end{align}

For the amplitude to adhere to uniform transcendentality \cite{Schlotterer:2012ny}, the expressions in the parentheses that multiply the subleading transcendental zeta-values in (\ref{eq: xisint4pt}), (\ref{eq: xipsint}), and (\ref{eq: xihpsint}) together with the integrals over $\mathcal{F}_R$ should have transcendental weight one. In \cite{Claasen:2024ssh}, we proposed that assigning transcendental weight one to $\frac{\zeta_n'}{\zeta_n}-n\gamma_E$ would revive the uniform transcendentality for the four-point amplitudes. However, in the case of five-point amplitudes, as we lack the data from the integrals over $\mathcal{F}_R$, we cannot say anything meaningful about transcendentality.

\section{Summary and Outlook}
In this paper, we have explored the low-energy expansion of massless five-point type IIB superstring amplitudes in ten-dimensional flat spacetime at one loop, reaching the tenth order in $\ap$ relative to the supergravity amplitude across all sectors. These results correspond to terms in the effective action up to $D^{12}R^5$ and $D^{14}\phi R^4$. Following the pure spinor formalism approach in \cite{Green:2013bza}, these amplitudes are expressed as integrals over the configuration space of four points on a torus, followed by an integration over the moduli space of tori. However, a direct low-energy expansion of the configuration space integral cannot serve as the integrand for the modular integral because the non-analytic structure of the amplitude leads to divergences. To address this, we employed the standard prescription of splitting the fundamental domain $\mathcal{F}$ into two regions, as shown in (\ref{eq:split}).

Our primary advancement involved converting the low-energy expansion of the configuration space integrals, known as modular graph forms (MGFs), into iterated integrals of (anti-)holomorphic modular forms within the formalism of equivariant iterated Eisenstein integrals \cite{Brown:2017qwo,Dorigoni:2022npe,Dorigoni:2024oft}. Using and extending the conversion algorithm from \cite{Claasen:2025vcd}, we accounted for the new graph topologies inherent to five-point MGFs. Expanding these integrals in terms of iterated integrals, denoted $\beta^{\text{eqv}}$, allowed us to map them onto a basis suitable for moduli space integration. This basis, introduced in \cite{Dorigoni:2021jfr,Dorigoni:2024oft}, is defined by Poisson equations and $\beta^{\text{eqv}}$ expansions.

While the integration of this basis was largely established up to depth three in \cite{Doroudiani:2023bfw}, we encountered a unique case that was absent in the four-point calculations. Although an analytical solution for this element remains elusive, we developed a technique (detailed in appendix \ref{sec:appendix}) to reduce the expression to a one-dimensional integral amenable to numerical evaluation. This machinery allowed us to systematically expand and integrate the MGFs over the lower part of the fundamental domain.

Non-vanishing five-point amplitudes emerge in two distinct sectors: one that conserves R-symmetry and another that maximally violates it. We presented the results for both using kinematic $M$-matrices, which are standard for tree-level amplitudes and universally capture open- and closed-string low-energy expansions. Beyond the five-point tree-level $M$-matrices that combined with four-point one-loop coefficients, new kinematic structures were needed to describe the five-point amplitude, signaling genuine non-factorizable five-point interactions. Furthermore, the inability to distinguish $\{M_3,M_7\}$ from $\{M_5,M_5\}$ at four points led to a one-parameter freedom in the presentation of the five-point amplitudes, which is a feature that is new to the results presented in this paper. Additionally, as noted in \cite{Boels:2013jua}, the maximally R-violating sector can be expanded in terms of symmetric polynomials of Mandelstam invariants; these results are provided in (\ref{eq:violboels}). We further related the amplitudes in these sectors to their corresponding modular forms in the type IIB effective action, and by $S$-duality studied the relationship between the two sectors.

The coefficients $\Xi_{\bullet}$ appearing in the perturbative terms for four- and five-point amplitudes are rational linear combinations of Riemann zeta values, the Euler-Mascheroni constant $\gamma_E$, the logarithmic derivatives of the Riemann zeta function at odd integers, and a novel numerical value $\omega$ of unknown transcendental nature. The reason indecomposable multiple zeta values beyond depth 1 do not appear in the result is that conjecturally, only the single-valued versions of them are expected to contribute, and the order in $\ap$ in this work is not high enough to see non-reducible multiple zeta values, where the first instance of them is at weight 11 (which corresponds to $\ap^{14}$). We emphasize that the $\Xi_{\bullet}$ result from integration over the lower part of the fundamental domain; consequently, the rational coefficients of these transcendental numbers are expected to shift upon inclusion of the upper part of the integral. However, for the four-point case where the upper integral is known \cite{DHoker:2019blr}, only the coefficients of $\gamma_E$ and the Riemann zeta value of lowest transcendental weight are modified. Notably, we observed a consistent pattern across all 18 calculated $\Xi_{\bullet}$ values: the sum of the coefficients of the Euler-Mascheroni constant and the logarithmic derivatives of the Riemann zeta functions is always equal to 1.

Our results open several promising avenues for future research. A primary next step is to combine the current findings with the integration over the upper part of the fundamental domain. This region cannot be addressed via the standard MGF expansion and requires an approach similar to that developed in \cite{DHoker:2019blr}. Furthermore, the non-analytic structure and discontinuities of the amplitude resulting from the upper-part integration could be studied independently using unitarity and factorization methods.

Due to the current lack of data and efficient techniques for calculating the constant terms in the zero-mode of the Fourier expansion of MGFs, we were unable to determine the rational coefficients for the Riemann zeta values at the highest transcendental weights. One potential strategy to extract these coefficients is to analyze the UV divergences in the one-loop matrix elements of the effective action, following the methodology in \cite{Edison:2021ebi}. Additionally, the effective action presented in this work remains schematic; for instance, the explicit form of the covariant derivatives has not been fully articulated. We leave the derivation of these explicit expressions for future work.

Another natural extension is the generalization of this framework to higher-point amplitudes and higher orders in the $\alpha'$ expansion. The latter will necessitate new integration techniques for iterated integrals of depth four and beyond. While the methods introduced in this paper allow for the numerical evaluation of such integrals, their systematic analytical calculation remains an open problem. Finally, the number-theoretic patterns emerging in these one-loop amplitudes merit further investigation, as such structures often provide deep insights into the underlying mathematical and physical principles of superstring theory.

\acknowledgments{We thank Oliver Schlotterer for valuable discussions at several stages of this work. We furthermore thank Axel Kleinschmidt and Oliver Schlotterer for their useful comments on the manuscript. We also thank the ICTP winter school on number theory and physics (smr 4162) and ICTP for hospitality. MD was supported by the ERE grant RF$\backslash$ERE$\backslash$221103 associated with the Royal Society University Research Fellowship Grant URF$\backslash$R1$\backslash$221236. The research of EC is funded by the European Union (ERC Synergy Grant MaScAmp 101167287). Views and opinions expressed are however those of the authors only and do not necessarily reflect those of the European Union or the European Research Council. Neither the European Union nor the granting authority can be held responsible for them.}

\appendix
\section{Five-point integrands}
\label{appendix: beqvs}
In this appendix, we collect all integral expressions relevant to the four- and five-point calculations in section \ref{sec:gen1}. We write them in terms of integrals of iterated Eisenstein integrals over the truncated fundamental domain of $\SL(2,\mathbb Z)$ defined in \eqref{eq:split}. All expressions can also be found in the ancillary file. At four points, the relevant integrals are 
\begin{align}
    \Xi_3&:= \int_{\mathcal{F}_L} \frac{d^2\tau}{\tau_2^2}\{2\pi\}\,, \nonumber\\
    \Xi_5&:= \int_{\mathcal{F}_L} \frac{d^2\tau}{\tau_2^2}\Big\{-24\pi \beqv{1}{4}\Big\}\,,\nonumber\\
    \Xi_{3,3}&:= \int_{\mathcal{F}_L} \frac{d^2\tau}{\tau_2^2}\Big\{(-2\pi)\Big(150\beqv{2}{6}-\zeta_3\Big)\Big\}\,,\nonumber\\
    \Xi_7&:= \int_{\mathcal{F}_L} \frac{d^2\tau}{\tau_2^2}\Big\{(-16\pi)\Big(28\beqv{3}{8}-18\beqv{1\,1}{4\,4}+9\beqv{2\,0}{4\,4}\Big)\Big\}\,,\nonumber\\
    \Xi_{\{3,5\}}&:= \int_{\mathcal{F}_L} \frac{d^2\tau}{\tau_2^2}\Big\{(-12\pi)\Big(189\beqv{4}{10}-150\beqv{1\,2}{4\,6}+140\beqv{2\,1}{4\,6}-150\beqv{2\,2}{6\,4}\nonumber\\&\qquad+140\beqv{3\,0}{6\,4}+\beqv{1}{4}\zeta_3\Big)+\frac{29\pi}{30}\zeta_5\Big\}\,,\nonumber\\
    \Xi_{9}&:=\int_{\mathcal{F}_L} \frac{d^2\tau}{\tau_2^2}\Big\{(-32\pi)\Bigl(737\beqv{5}{12}-168\beqv{1\,3}{4\,8}+252\beqv{2\,2}{4\,8}+200\beqv{3\,1}{6\,6}\nonumber\\&\qquad-168\beqv{3\,1}{8\,4}+100\beqv{4\,0}{6\,6}+252\beqv{4\,0}{8\,4}+108\beqv{1\,1\,1}{4\,4\,4}-54\beqv{1\,2\,0}{4\,4\,4}\nonumber\\&\qquad-54\beqv{2\,0\,1}{4\,4\,4}\Big)\Big\}\,,\nonumber\\    
    \Xi_{3,3,3}&:=\int_{\mathcal{F}_L} \frac{d^2\tau}{\tau_2^2}\Big\{\Big(\frac{\pi}{3}\Big)\Big(252472\beqv{5}{12}+100400\beqv{4\,0}{6\,6}+26208\beqv{2\,2}{4\,8}+135000\beqv{2\,2}{6\,6}\nonumber\\&\qquad-5376\beqv{1\,3}{4\,8}-94400\beqv{3\,1}{6\,6}-5376\beqv{3\,1}{8\,4}+26208\beqv{4\,0}{8\,4}+3456\beqv{1\,1\,1}{4\,4\,4}\nonumber\\&\qquad-1728\beqv{1\,2\,0}{4\,4\,4}-1728\beqv{2\,0\,1}{4\,4\,4}+3888\beqv{2\,1\,0}{4\,4\,4}-900\beqv{2}{6}\zeta_3+3\zeta_3^2\Big)\Big\}\,,\nonumber\\
    \Xi_{\{3,7\}}&:=\int_{\mathcal{F}_L} \frac{d^2\tau}{\tau_2^2}\Big\{(12\pi)\Big(-\tfrac{29588}{3}\beqv{6}{14}+2268\beqv{1\,4}{4\,10}-3456\beqv{2\,3}{4\,10}+2800\beqv{2\,3}{6\,8}\nonumber\\&\qquad-6720\beqv{3\,2}{6\,8}+2800\beqv{3\,2}{8\,6}+1540\beqv{4\,1}{6\,8}-6720\beqv{4\,1}{8\,6}+2268\beqv{4\,1}{10\,4}\nonumber\\&\qquad+1540\beqv{5\,0}
    {8\,6}-3456\beqv{5\,0}{10\,4}-1800\beqv{1\,1\,2}{4\,4\,6}+1680\beqv{1\,2\,1}{4\,4\,6}-1800\beqv{1\,2\,1}{4\,6\,4}\nonumber\\&\qquad+1680\beqv{1\,3\,0}{4\,6\,4}+900\beqv{2\,0\,2}{4\,4\,6}+240\beqv{2\,1\,1}{4\,4\,6}+1680\beqv{2\,1\,1}{4\,6\,4}-1800\beqv{2\,1\,1}{6\,4\,4}\nonumber\\&\qquad-720\beqv{2\,2\,0}{4\,4\,6}-1260\beqv{2\,2\,0}{4\,6\,4}+900\beqv{2\,2\,0}{6\,4\,4}+1680\beqv{3\,0\,1}{6\,4\,4}+240\beqv{3\,1\,0}{6\,4\,4}\nonumber\\&\qquad-720\beqv{4\,0\,0}{6\,4\,4}-\tfrac{56}{3}\beqv{3}{8}\zeta_3+12\beqv{1\,1}{4\,4}\zeta_3-6\beqv{2\,0}{4\,4}\zeta_3-\tfrac{29}{30}\beqv{1}{4}\zeta_5\nonumber\\&\qquad-\tfrac{163}{30240}\zeta_7\Big)\Big\}\,.
    \label{eq: xis4pt}
\end{align}
The additional integrals relevant to the five-point amplitude in the conserving sector read
\begin{align}
    \Xi_{7^\prime}&:=\int_{\mathcal{F}_L} \frac{{\rm d}^2\tau}{\tau_2^2}\Big\{(-15\pi)\Big(196\beqv{3}{8}+30\beqv{1\,1}{4\,4}+9\beqv{2\,0}{4\,4}\Big)\Big\}\,,\nonumber\\
    \Xi_{8^\prime}&:=\int_{\mathcal{F}_L} \frac{{\rm d}^2\tau}{\tau_2^2}\Big\{\Big(\frac{36\pi}{5}\Big)\Big(2307\beqv{4}{10}+150\beqv{1\,2}{4\,6}+260\beqv{2\,1}{4\,6}+150\beqv{2\,1}{6\,4}\nonumber\\&\qquad+260\beqv{3\,0}{6\,4}-\beqv{1}{4}\zeta_3-\tfrac{7}{360}\zeta_5\Big)\Big\}\,,\nonumber\\
    \Xi_{9^\prime}&:=\int_{\mathcal{F}_L} \frac{{\rm d}^2\tau}{\tau_2^2}\Big\{\Big(-\frac{2\pi}{9}\Big)\Big(2402818\beqv{5}{12}-10920\beqv{1\,3}{4\,8}+394002\beqv{2\,2}{4\,8}\nonumber\\&\qquad+50625\beqv{2\,2}{6\,6}+424300\beqv{3\,1}{6\,6}-10920\beqv{3\,1}{8\,4}+102575\beqv{4\,0}{6\,6}\nonumber\\&\qquad+18(21889\beqv{4\,0}{8\,4}-2148\beqv{1\,1\,1}{4\,4\,4}+750\beqv{1\,2\,0}{4\,4\,4}+750\beqv{2\,0\,1}{4\,4\,4}\nonumber\\&\qquad+1863\beqv{2\,1\,0}{4\,4\,4})-\tfrac{675}{2}\beqv{2}{6}\zeta_3\Big)+\pi c_{9^\prime}\zeta_3^2\Big\}\,,\nonumber\\
    \Xi_{9^{\prime\prime}}&:=\int_{\mathcal{F}_L} \frac{{\rm d}^2\tau}{\tau_2^2}\Big\{\Big(-\frac{\pi}{9}\Big)\Big(1295987\betaeqv{5\\12}+ 2\big(18564\betaeqv{1&3\\4&8}+ 76482\betaeqv{2&2\\4&8} \nonumber\\&\qquad+ 160600\betaeqv{3&1\\6&6}+ 18564\betaeqv{3&1\\8&4}+ 40025\betaeqv{4&0\\6&6}+ 76482\betaeqv{4&0\\8&4} \nonumber\\&\qquad- 3186\betaeqv{1&1&1\\4&4&4}+ 3537\betaeqv{1&2&0\\4&4&4}+ 3537\betaeqv{2&0&1\\4&4&4}+ 7776\betaeqv{2&1&0\\4&4&4}\big)\Big)\nonumber\\&\qquad+\pi c_{9^{\prime\prime}}\zeta_3^2\Big\}\,,\nonumber\\
    \Xi_{\{3,7'\}}&:=\int_{\mathcal{F}_L}\frac{{\rm d}^2\tau}{\tau_2^2}\Big\{\Big(\frac{360\pi}{28921296}\Big)\Big(59848146318\betaeqv{1&4\\4&10}
+ 21069993813\betaeqv{2&3\\4&10} \nonumber\\
&\qquad + 1661511250\betaeqv{3&2\\6&8}
- 160105744995\betaeqv{4&1\\6&8}
+ 1661511250\betaeqv{4&1\\8&6} \nonumber\\
&\qquad + 59848146318\betaeqv{4&1\\10&4}
- 160105744995\betaeqv{5&0\\8&6}
+ 21069993813\betaeqv{5&0\\10&4} \nonumber\\
&\qquad + 5\big(
927412410\betaeqv{1&1&2\\4&4&6}
+ 162326260\betaeqv{1&2&1\\4&4&6}
+ 1003436400\betaeqv{1&2&1\\4&6&4} \nonumber\\
&\qquad + 922205674\betaeqv{1&3&0\\4&6&4}
- 186778425\betaeqv{2&0&2\\4&4&6}
+ 591742560\betaeqv{2&1&1\\4&4&6} \nonumber\\
&\qquad + 922205674\betaeqv{2&1&1\\4&6&4}
+ 927412410\betaeqv{2&1&1\\6&4&4}
- 1602021680\betaeqv{2&2&0\\4&4&6} \nonumber\\
&\qquad + 2483450748\betaeqv{2&2&0\\4&6&4}
- 186778425\betaeqv{2&2&0\\6&4&4}
+ 20\big(
8116313\betaeqv{3&0&1\\6&4&4} \nonumber\\
&\qquad + 29587128\betaeqv{3&1&0\\6&4&4}
- 80101084\betaeqv{4&0&0\\6&4&4}
\big)\big)-30913747\betaeqv{1&1\\4&4}\zeta_3 \nonumber\\
&\qquad
+ \tfrac{12451895}{2}\betaeqv{2&0\\4&4}\zeta_3
\big) - \tfrac{2095009183}{360}\betaeqv{1\\4}\zeta_5
\Big)+\pi c_{\{3,7'\}}\zeta_7\Big\}
\nonumber\,,\\
    \Xi_{10^{\prime}}&:=
\int_{\mathcal{F}_L} \frac{{\rm d}^2\tau}{\tau_2^2}\Big\{\Big(-\frac{\pi}{18075810}\Big)\Big(
795118354920 \betaeqv{1&4\\4&10}
- 955112585130 \betaeqv{2&3\\4&10}
\nonumber\\
&\qquad + 154828825200 \betaeqv{3&2\\6&8}
+ 10839392550 \betaeqv{4&1\\6&8}
+ 154828825200 \betaeqv{4&1\\8&6}
\nonumber\\
&\qquad + 795118354920 \betaeqv{4&1\\10&4}
+ 10839392550 \betaeqv{5&0\\8&6}
- 955112585130 \betaeqv{5&0\\10&4}
\nonumber\\
&\qquad + 83775816000 \betaeqv{1&1&2\\4&4&6}
+ 65496150000 \betaeqv{1&2&1\\4&4&6}
+ 567654944400 \betaeqv{1&2&1\\4&6&4}
\nonumber\\
&\qquad - 289584295200 \betaeqv{1&3&0\\4&6&4}
- 32626260000 \betaeqv{2&0&2\\4&4&6}
- 40652409600 \betaeqv{2&1&1\\4&4&6}
\nonumber\\
&\qquad - 289584295200 \betaeqv{2&1&1\\4&6&4}
+ 83775816000 \betaeqv{2&1&1\\6&4&4}
- 32927121450 \betaeqv{2&2&0\\4&4&6}
\nonumber\\
&\qquad + 79337070000 \betaeqv{2&2&0\\4&6&4}
- 32626260000 \betaeqv{2&2&0\\6&4&4}
+ 65496150000 \betaeqv{3&0&1\\6&4&4}
\nonumber\\
&\qquad - 40652409600 \betaeqv{3&1&0\\6&4&4}
- 32927121450 \betaeqv{4&0&0\\6&4&4}
- 558505440 \betaeqv{1&1\\4&4}\zeta_3
\nonumber\\
&\qquad + 217508400 \betaeqv{2&0\\4&4}\zeta_3
+ 122998753 \betaeqv{1\\4}\zeta_5
\Big)+\pi c_{10'}\zeta_7\Big\}\,,\nonumber\\
    \Xi_{10^{\prime\prime}}&:=
\int_{\mathcal{F}_L}\frac{{\rm d}^2\tau}{\tau_2^2}\Big\{\Big(\frac{\pi}{126530670}\Big)\Big(
4930632540 \betaeqv{1&4\\4&10}
+ 102662393625 \betaeqv{2&3\\4&10}
\nonumber\\
&\qquad - 18979600500 \betaeqv{2&3\\6&8}
+ 19128501000 \betaeqv{3&2\\6&8}
- 18979600500 \betaeqv{3&2\\8&6}
\nonumber\\
&\qquad - 210566886075 \betaeqv{4&1\\6&8}
+ 19128501000 \betaeqv{4&1\\8&6}
+ 4930632540 \betaeqv{4&1\\10&4}
\nonumber\\
&\qquad - 210566886075 \betaeqv{5&0\\8&6}
+ 102662393625 \betaeqv{5&0\\10&4}
- 4045437000 \betaeqv{1&1&2\\4&4&6}
\nonumber\\
&\qquad - 4023617400 \betaeqv{1&2&1\\4&4&6}
- 4457305800 \betaeqv{1&2&1\\4&6&4}
+ 3156795000 \betaeqv{1&3&0\\4&6&4}
\nonumber\\
&\qquad + 226611000 \betaeqv{2&0&2\\4&4&6}
+ 5968911600 \betaeqv{2&1&1\\4&4&6}
+ 3156795000 \betaeqv{2&1&1\\4&6&4}
\nonumber\\
&\qquad - 4045437000 \betaeqv{2&1&1\\6&4&4}
- 8615499975 \betaeqv{2&2&0\\4&4&6}
+ 24104611800 \betaeqv{2&2&0\\4&6&4}
\nonumber\\
&\qquad + 226611000 \betaeqv{2&2&0\\6&4&4}
- 4023617400 \betaeqv{3&0&1\\6&4&4}
+ 5968911600 \betaeqv{3&1&0\\6&4&4}
\nonumber\\
&\qquad - 8615499975 \betaeqv{4&0&0\\6&4&4}
+ 126530670 \betaeqv{3\\8}\zeta_3
+ 26969580 \betaeqv{1&1\\4&4}\zeta_3
\nonumber\\
&\qquad - 1510740 \betaeqv{2&0\\4&4}\zeta_3
- 8354324 \betaeqv{1\\4}\zeta_5
\Big)+\pi c_{10''}\zeta_7\Big\}\,,\nonumber\\
    \Xi_{10^{\prime\prime\prime}}&:=\int_{\mathcal{F}_L} \frac{{\rm d}^2\tau}{\tau_2^2}\Big\{\Big(\frac{\pi}{3759768480}\Big)\Big(
3759768480 \betaeqv{6\\14}
- 47225160 \betaeqv{1&4\\4&10}
\nonumber\\
&\qquad + 689627250 \betaeqv{2&3\\4&10}
+ 634158000 \betaeqv{3&2\\6&8}
+ 85349250 \betaeqv{4&1\\6&8}
\nonumber\\
&\qquad + 634158000 \betaeqv{4&1\\8&6}
- 47225160 \betaeqv{4&1\\10&4}
+ 85349250 \betaeqv{5&0\\8&6}
\nonumber\\
&\qquad + 689627250 \betaeqv{5&0\\10&4}
- 21492000 \betaeqv{1&1&2\\4&4&6}
+ 4899600 \betaeqv{1&2&1\\4&4&6}
\nonumber\\
&\qquad - 20422800 \betaeqv{1&2&1\\4&6&4}
+ 4176000 \betaeqv{1&3&0\\4&6&4}
+ 8370000 \betaeqv{2&0&2\\4&4&6}
\nonumber\\
&\qquad + 63849600 \betaeqv{2&1&1\\4&4&6}
+ 4176000 \betaeqv{2&1&1\\4&6&4}
- 21492000 \betaeqv{2&1&1\\6&4&4}
\nonumber\\
&\qquad + 10325250 \betaeqv{2&2&0\\4&4&6}
+ 94834800 \betaeqv{2&2&0\\4&6&4}
+ 8370000 \betaeqv{2&2&0\\6&4&4}
\nonumber\\
&\qquad + 4899600 \betaeqv{3&0&1\\6&4&4}
+ 63849600 \betaeqv{3&1&0\\6&4&4}
+ 10325250 \betaeqv{4&0&0\\6&4&4}
\nonumber\\
&\qquad + 143280 \betaeqv{1&1\\4&4}\zeta_3
- 55800 \betaeqv{2&0\\4&4}\zeta_3
- 17179 \betaeqv{1\\4}\zeta_5
\Big)+\pi c_{10'''}\zeta_7\Big\}\,.
    \label{eq: xips}
\end{align}
Note that results up to $\ap^9$ were already found in \cite{Green:2013bza}. However, direct translation of their results into the iterated integral formalism does not coincide with the results presented here for the following reason. The authors of \cite{Green:2013bza} worked out the relations between the modular graph forms that appear in the amplitude to simplify the presentation, but were only able to find relations between the real parts of the modular graph forms. As they assumed that there would be no imaginary parts involved, this led to incomplete relations. In fact, imaginary cusp form MGFs start to enter the calculations from weight $\ap^8$ onward, and arise in such a way that the final integrand is real, as is the case in the expressions (\ref{eq: xips}). This is not the case in \cite{Green:2013bza}, in which the integrands in eqns. (5.6) and (5.8) turn out to be complex. In the end, this does not matter for the amplitudes, as the real parts agree and the integral over any imaginary cusp form MGF can be shown to vanish \cite{Doroudiani:2023bfw}. 

Finally, the integrals in the violating sector yield
\begin{align}
    \hat{\Xi}_{7^\prime}&:=\int_{\mathcal{F}_L} \frac{{\rm d}^2\tau}{\tau_2^2} \Big\{ (-3\pi)\Big(1988\betaeqv{3\\8}+366\betaeqv{1&1\\4&4}+129\betaeqv{2&0\\4&4}\Big)\Big\} \,,\nonumber\\
    \hat{\Xi}_{8^\prime}&:= \int_{\mathcal{F}_L} \frac{{\rm d}^2\tau}{\tau_2^2}\Big\{\Big(\frac{18\pi}{5}\Big)\Big( 6663\betaeqv{4\\10}+ 750\betaeqv{1&2\\4&6}
+ 180\betaeqv{2&1\\4&6}+ 750\betaeqv{2&1\\6&4} \nonumber\\
&\qquad + 180\betaeqv{3&0\\6&4} - 5\betaeqv{1\\4}\zeta_3
\Big) -\tfrac{303}{100}\zeta_5\Big\}\,,\nonumber\\
    \hat{\Xi}_{9^\prime}&:=\int_{\mathcal{F}_L} \frac{{\rm d}^2\tau}{\tau_2^2}\Big\{\Big(-\frac{\pi}{81}\Big)\Big(
49850548 \betaeqv{5\\12}
- 2467920 \betaeqv{1&3\\4&8}+ 10254132 \betaeqv{2&2\\4&8}
\nonumber\\
&\qquad 
- 303750 \betaeqv{2&2\\6&6}
+ 10017400 \betaeqv{3&1\\6&6}- 2467920 \betaeqv{3&1\\8&4}
+ 4737350 \betaeqv{4&0\\6&6}
\nonumber\\
&\qquad 
+ 10254132 \betaeqv{4&0\\8&4}
- 1755216 \betaeqv{1&1&1\\4&4&4}
+ 212760 \betaeqv{1&2&0\\4&4&4}
\nonumber\\
&\qquad + 212760 \betaeqv{2&0&1\\4&4&4} + 638604 \betaeqv{2&1&0\\4&4&4}
+ 2025 \betaeqv{2\\6}\zeta_3\Big)
+ \pi \hat{c}_{9'}\zeta_3^2
\Big\}\,,\nonumber\\
    \hat{\Xi}_{9^{\prime\prime}}&:=\int_{\mathcal{F}_L} \frac{{\rm d}^2\tau}{\tau_2^2}\Big\{\Big(-\frac{\pi}{81}\Big)\Big(
10557151 \betaeqv{5\\12}
- 394968 \betaeqv{1&3\\4&8}+ 1898820 \betaeqv{2&2\\4&8}
\nonumber\\
&\qquad 
- 1215000 \betaeqv{2&2\\6&6}
+ 4795600 \betaeqv{3&1\\6&6} - 394968 \betaeqv{3&1\\8&4}
+ 1089650 \betaeqv{4&0\\6&6}
\nonumber\\
&\qquad
+ 1898820 \betaeqv{4&0\\8&4}- 218484 \betaeqv{1&1&1\\4&4&4}
+ 4266 \betaeqv{1&2&0\\4&4&4}
+ 4266 \betaeqv{2&0&1\\4&4&4}
\nonumber\\
&\qquad - 244944 \betaeqv{2&1&0\\4&4&4}
+ 8100 \betaeqv{2\\6}\zeta_3
\Big)
+\pi\hat{c}_{9''}\zeta_3^2\Big\}\,,\nonumber\\
    \hat{\Xi}_{9^{\prime\prime\prime}}&:=\int_{\mathcal{F}_L} \frac{{\rm d}^2\tau}{\tau_2^2}\Big\{\Big(\frac{64\pi}{3}\Big)\Big(
 1573\betaeqv{5\\12}
- 336\betaeqv{1&3\\4&8}  + 1071\betaeqv{2&2\\4&8}
+ 400\betaeqv{3&1\\6&6}\nonumber\\
&\qquad
- 336\betaeqv{3&1\\8&4}  - 25\betaeqv{4&0\\6&6}
+ 1071\betaeqv{4&0\\8&4}
+ 216\betaeqv{1&1&1\\4&4&4} \nonumber\\
&\qquad - 108
\betaeqv{1&2&0\\4&4&4}
- 108\betaeqv{2&0&1\\4&4&4}
- 486\betaeqv{2&1&0\\4&4&4}\Big)
+\pi\hat{c}_{9'''}\zeta_3^2\Big\}\,,\nonumber\\
    \hat{\Xi}_{\{3,7'\}}&:=
\int_{\mathcal{F}_L} \frac{{\rm d}^2\tau}{\tau_2^2}\Big\{\Big(-\frac{\pi}{967600}\Big)\Big(
210703088960 \betaeqv{6\\14}
- 732208124880 \betaeqv{1&4\\4&10}
\nonumber\\
&\qquad + 908360046360 \betaeqv{2&3\\4&10}
- 383526192000 \betaeqv{2&3\\6&8}
+ 494246768400 \betaeqv{3&2\\6&8}
\nonumber\\
&\qquad - 383526192000 \betaeqv{3&2\\8&6}
- 406910250600 \betaeqv{4&1\\6&8}
+ 494246768400 \betaeqv{4&1\\8&6}
\nonumber\\
&\qquad - 732208124880 \betaeqv{4&1\\10&4}
- 406910250600 \betaeqv{5&0\\8&6}
+ 908360046360 \betaeqv{5&0\\10&4}
\nonumber\\
&\qquad - 165572478000 \betaeqv{1&1&2\\4&4&6}
+ 7375456800 \betaeqv{1&2&1\\4&4&6}
- 174414124800 \betaeqv{1&2&1\\4&6&4}
\nonumber\\
&\qquad + 35941093200 \betaeqv{1&3&0\\4&6&4}
+ 5716899000 \betaeqv{2&0&2\\4&4&6}
+ 80432265600 \betaeqv{2&1&1\\4&4&6}
\nonumber\\
&\qquad + 35941093200 \betaeqv{2&1&1\\4&6&4}
- 165572478000 \betaeqv{2&1&1\\6&4&4}
- 65998206000 \betaeqv{2&2&0\\4&4&6}
\nonumber\\
&\qquad + 26189892000 \betaeqv{2&2&0\\4&6&4}
+ 5716899000 \betaeqv{2&2&0\\6&4&4}
+ 7375456800 \betaeqv{3&0&1\\6&4&4}
\nonumber\\
&\qquad + 80432265600 \betaeqv{3&1&0\\6&4&4}
- 65998206000 \betaeqv{4&0&0\\6&4&4}
+ 2556841280 \betaeqv{3\\8}\zeta_3
\nonumber\\
&\qquad + 1103816520 \betaeqv{1&1\\4&4}\zeta_3
- 38112660 \betaeqv{2&0\\4&4}\zeta_3
+ 6228793 \betaeqv{1\\4}\zeta_5
\Big)+\pi \hat{c}_{\{3,7'\}}\zeta_7
\Big\}\,,\nonumber\\
    \hat{\Xi}_{10^{\prime}}&:=
\int_{\mathcal{F}_L} \frac{{\rm d}^2\tau}{\tau_2^2}\Big\{\Big(\frac{\pi}{120950}\Big)\Big(
16304181920 \betaeqv{6\\14}
- 8742402360 \betaeqv{1&4\\4&10}
\nonumber\\
&\qquad + 4020824070 \betaeqv{2&3\\4&10}
- 3476424000 \betaeqv{2&3\\6&8}
+ 11698486800 \betaeqv{3&2\\6&8}
\nonumber\\
&\qquad - 3476424000 \betaeqv{3&2\\8&6}
+ 3267464550 \betaeqv{4&1\\6&8}
+ 11698486800 \betaeqv{4&1\\8&6}
\nonumber\\
&\qquad - 8742402360 \betaeqv{4&1\\10&4}
+ 3267464550 \betaeqv{5&0\\8&6}
+ 4020824070 \betaeqv{5&0\\10&4}
\nonumber\\
&\qquad + 2234844000 \betaeqv{1&1&2\\4&4&6}
- 2775308400 \betaeqv{1&2&1\\4&4&6}
- 988707600 \betaeqv{1&2&1\\4&6&4}
\nonumber\\
&\qquad - 647121600 \betaeqv{1&3&0\\4&6&4}
- 1117422000 \betaeqv{2&0&2\\4&4&6}
- 3397420800 \betaeqv{2&1&1\\4&4&6}
\nonumber\\
&\qquad - 647121600 \betaeqv{2&1&1\\4&6&4}
+ 2234844000 \betaeqv{2&1&1\\6&4&4}
+ 2139873750 \betaeqv{2&2&0\\4&4&6}
\nonumber\\
&\qquad - 2720790000 \betaeqv{2&2&0\\4&6&4}
- 1117422000 \betaeqv{2&2&0\\6&4&4}
- 2775308400 \betaeqv{3&0&1\\6&4&4}
\nonumber\\
&\qquad - 3397420800 \betaeqv{3&1&0\\6&4&4}
+ 2139873750 \betaeqv{4&0&0\\6&4&4}
+ 23176160 \betaeqv{3\\8}\zeta_3
\nonumber\\
&\qquad - 14898960 \betaeqv{1&1\\4&4}\zeta_3
+ 7449480 \betaeqv{2&0\\4&4}\zeta_3
+ 853821 \betaeqv{1\\4}\zeta_5
\Big)+\pi \hat{c}_{10'}\zeta_7
\Big\}\,,\nonumber\\
    \hat{\Xi}_{10^{\prime\prime\prime}}&:=
\int_{\mathcal{F}_L} \frac{{\rm d}^2\tau}{\tau_2^2}\Big\{\Big(\frac{\pi}{25157600}\Big)\Big(
31102240 \betaeqv{6\\14}
- 657720 \betaeqv{1&4\\4&10}+ 5490990 \betaeqv{2&3\\4&10}
\nonumber\\
&\qquad 
+ 672000 \betaeqv{2&3\\6&8}
+ 3183600 \betaeqv{3&2\\6&8}+ 672000 \betaeqv{3&2\\8&6}
+ 4438350 \betaeqv{4&1\\6&8}
\nonumber\\
&\qquad 
+ 3183600 \betaeqv{4&1\\8&6}
- 657720 \betaeqv{4&1\\10&4}
+ 4438350 \betaeqv{5&0\\8&6}
+ 5490990 \betaeqv{5&0\\10&4}
\nonumber\\
&\qquad - 432000 \betaeqv{1&1&2\\4&4&6}
+ 349200 \betaeqv{1&2&1\\4&4&6}
- 421200 \betaeqv{1&2&1\\4&6&4}+ 280800 \betaeqv{1&3&0\\4&6&4}
\nonumber\\
&\qquad 
+ 216000 \betaeqv{2&0&2\\4&4&6}
+ 1094400 \betaeqv{2&1&1\\4&4&6}+ 280800 \betaeqv{2&1&1\\4&6&4}
- 432000 \betaeqv{2&1&1\\6&4&4}
\nonumber\\
&\qquad 
- 47250 \betaeqv{2&2&0\\4&4&6}
+ 810000 \betaeqv{2&2&0\\4&6&4}
+ 216000 \betaeqv{2&2&0\\6&4&4}
+ 349200 \betaeqv{3&0&1\\6&4&4}
\nonumber\\
&\qquad + 1094400 \betaeqv{3&1&0\\6&4&4}
- 47250 \betaeqv{4&0&0\\6&4&4}
- 4480 \betaeqv{3\\8}\zeta_3+ 2880 \betaeqv{1&1\\4&4}\zeta_3
\nonumber\\
&\qquad 
- 1440 \betaeqv{2&0\\4&4}\zeta_3
+ 67 \betaeqv{1\\4}\zeta_5
\Big)+\pi\hat{c}_{10'''}\zeta_7
\Big\}\,.
\label{eq: xihps}
\end{align}
Similar considerations regarding mismatches due to cusp forms with \cite{Green:2013bza} apply to these integrals.

\section{Numerical integration technique}
\label{sec:appendix}
In the calculation of genus-one string amplitudes in flat ten-dimensional spacetime, we encounter integrals over the complex structure modulus
\begin{align}
    \int_{\mathcal{F}_L}\frac{d^2\tau}{\tau_2^2}F(\tau,\Bar{\tau})\,,
    \label{appB:int}
\end{align}
where $F(\tau,\Bar{\tau})$ is a general (non-)holomorphic modular function. Depending on the nature of $F(\tau,\Bar{\tau})$, different analytic evaluation techniques have been developed \cite{Doroudiani:2023bfw}, which encompass a wide range of functions $F(\tau,\Bar{\tau})$. Unfortunately, these methods are not sufficient to determine all integrals of interest in this work. Both as a sanity check and to push beyond what is analytically known, it would be helpful to get a handle on these integrals numerically. Such numerical methods have been developed in \cite{Manschot:2024prc}. However, we found a method that is much more practical and that requires only knowledge of the $q$-expansion of $F(\tau,\Bar{\tau})$. 
The idea is as follows. Given any (non-)holomorphic modular function $F(\tau,\Bar{\tau})$ with known $q$-expansion 
\begin{align}
    F(\tau,\Bar{\tau})=\sum_{m,n=0}^\infty \sum_{p}a_{p,m,n}\tau_2^p q^m \Bar{q}^n\, ,
\end{align}
where $p$ in the inner sum runs over a finite set of integers. This expands the integral (\ref{appB:int}) in a series of integrals. The idea now is to use Stokes theorem by rewriting each integrand as the anti-holomorphic derivative of another function. In other words, given a term $\tau_2^p q^m \Bar{q}^n \subset F(\tau,\Bar{\tau})$, we write its integral as
\begin{align}
    \int_{\mathcal{F}_L}\frac{d^2\tau}{\tau_2^2}\tau_2^p q^m \Bar{q}^n=\int_{\mathcal{F}_L}\frac{d^2\tau}{\tau_2^2}\overline{\nabla} \Tilde{F}_{p,m,n}=\int_{\partial\mathcal{F}_L} \Tilde{F}_{p,m,n}\,.
\end{align}
The trick now is to find a solution of $\Tilde{F}_{p,m,n}$ that is $T$-invariant so that the boundary integral simplifies to just an integral over an arc and a line. Writing the boundary integral as the union of
\begin{align}
    \partial\mathcal{F}_L = C_1 \cup C_2 \cup C_3 \cup C_4 \qquad 
    \begin{tikzpicture}[thick, >=stealth, baseline=(current bounding box.center), scale=0.7, every node/.style={transform shape}]
        \coordinate (BottomLeft) at (-1/2, {sqrt(3)/2});
        \coordinate (TopLeft) at (-1/2, 2);
        \coordinate (TopRight) at (1/2, 2);
        \coordinate (BottomRight) at (1/2, {sqrt(3)/2});
        \draw[->] (-1.5, 0) -- (1.75, 0); 
        \draw[->] (0, 0) -- (0, 3);      
        \draw[dashed, thin] (1,0) arc (0:180:1);
        \draw (BottomLeft) -- (TopLeft) node[midway, left] {$C_1$};
        \draw (-1/2,2) -- (0,2) node[midway, above] {$C_2$};
        \draw (0,2) -- (1/2,2);
        \draw (TopRight) -- (BottomRight) node[midway, right] {$C_3$};
        \draw (BottomRight) arc (60:120:1) node[midway, xshift=0.3cm, yshift=-0.3cm] {$C_4$};
        \foreach \x/\label in {-1/-1, -0.5/-\frac{1}{2}, 0.5/\frac{1}{2}, 1/1} {
            \draw (\x, 2pt) -- (\x, -2pt) node[below, font=\footnotesize] {$\label$};
        }
    \end{tikzpicture} \, .
\end{align}
By $T$-invariance, the integral over $C_1$ and $C_3$ cancel each other. Moreover, one can choose $\tilde{F}_{p,m,n}$ in a way that the integral over $C_2$ only contributes to the logarithmic and polynomial divergent terms, which can be calculated analytically and all of the numerical data of the integral (\ref{appB:int}) can be estimated by the numerical evaluation of the integral over the arc $C_4$. Such functions $\tilde{F}_{p,m,n}$ are not unique and we propose the following function
\begin{align}
    \tilde{F}(\tau,\Bar{\tau}) &= a_{-1,0,0}\log(\tau_2)+\sum_{p \neq -1} a_{p,0,0} \frac{\tau_2^{p-1}}{p-1}\notag \\
    &\quad -\sum_{\substack{m,n=0\\m+n\neq0}}^{\infty} \sum_{p\neq 0,1}a_{p,m,n}\tau_2^{p-1}e^{4\pi n\tau_2}E_{2-p}(4n\pi\tau_2)q^{m}\Bar{q}^{n} \, ,
    \label{Ftilde}
\end{align}
where $E_n(x)$ is the generalized exponential integral function
\begin{align}
    E_n(x)=\int_1^{\infty} e^{-x t} t^{-n} dt \, .
\end{align}
The expansion (\ref{Ftilde}) has all the properties we want, namely, it is $T$-invariant so its integral over $C_1$ and $C_3$ cancel each other. Also, the integral of its second line over $C_2$ vanishes so we have
\begin{align}
    \int_{C_2}\tilde{F}(\tau_1,L)\, d\tau_1 = a_{-1,0,0}\log(L)+\sum_{p\neq -1}\frac{L^{p-1}}{p-1}\, ,
\end{align}
which contains the divergent terms of the integral. By a simple analysis, one can see that for large $n$, we have
\begin{align}
    |E_{2-p}(4n\pi \tau_2)|< \frac{e^{-4n\pi\tau_2}}{4n\pi\tau_2-|p-2|} \, ,
\end{align}
and for large $m$ the integral is exponentially convergent because of $|q^m|=e^{4\pi m\tau_2}$. So we conclude that the integral $\int_{C_4}a_{p,m,n}\tau_2^{p-1}e^{4\pi n\tau_2}E_{2-p}(4n\pi\tau_2)q^{m}\Bar{q}^{n}$ exponentially decays by increasing $n$ and $m$. For this reason, this proposal is useful for high accuracy at rather low cutoffs for $n$ and $m$.
\section{Symmetric polynomials}
\label{appendix: symPol}
In this appendix, we gather the projections $\mathcal{P}$ defined in section \ref{ssec:viol} applied to all kinematic matrices $M_\bullet$ and its primed variants. We write the result in terms of the symmetric polynomials $\tau_k$ introduced in the same section \ref{ssec:viol}. We find
\begin{equation}
\begin{aligned}
    &\mathcal{P}(M_3)=3\,,\nonumber\\
    &\mathcal{P}(M_5)=\frac{5}{2}\tau_2\,,\nonumber\\
    &\mathcal{P}(M_3^2)=-\tau_3\,,\nonumber\\
    &\mathcal{P}(M_7)=\frac{21}{6}\tau_4+\frac{7}{64}\tau_2^2\,,\nonumber\\
    &\mathcal{P}(M_7')=-\frac{1}{3}\tau_4+\frac{1}{12}\tau_2^2\,,\nonumber\\
    &\mathcal{P}(\{M_3,M_5\})=-\frac{3}{5}\tau_5-\frac{5}{12}\tau_2\tau_3\,,\nonumber\\
    &\mathcal{P}(M_8')=-\frac{1}{3}\tau_5-\frac{5}{36}\tau_2\tau_3\,,\nonumber\\
\end{aligned}
\qquad
\begin{aligned}
    &\mathcal{P}(M_3^3)=\frac{1}{2}\tau_6+\frac{1}{2}\tau_3^2+\frac{1}{16}\tau_2^3-\frac{5}{8}\tau_2\tau_4-\frac{1}{2}\tau_6'\,,\nonumber\\
    &\mathcal{P}(M_9')=-\frac{2}{9}\tau_6-\frac{1}{9}\tau_3^2+\frac{1}{6}\tau_2\tau_4+\frac{2}{9}\tau_6'\,,\nonumber\\
    &\mathcal{P}(M_9'')=\frac{8}{9}\tau_6+\frac{4}{9}\tau_3^2+\frac{1}{12}\tau_2^3-\tau_2\tau_4-\frac{8}{9}\tau_6'\,,\nonumber\\
    &\mathcal{P}(M_9''')=\frac{2}{9}\tau_6+\frac{2}{9}\tau_3^2+\frac{1}{24}\tau_2^3-\frac{5}{12}\tau_2\tau_4-\frac{7}{9}\tau_6'\,,\nonumber\\
    &\mathcal{P}(\{M_3,M_7\})=-\frac{3}{7}\tau_7-\frac{7}{16}\tau_3\tau_4-\frac{7}{192}\tau_3\tau_2^2\,,\nonumber\\
    &\mathcal{P}(\{M_5,M_5\})=-\frac{1}{2}\tau_2\tau_5\,,\nonumber\\
    &\mathcal{P}(\{M_3,M_7'\})=\frac{1}{9}\tau_3\tau_4-\frac{1}{36}\tau_3\tau_2^2\,,\nonumber\\
\end{aligned}
\end{equation}
\vspace{-2em}
\begin{align}
    &\mathcal{P}(M_9)=\frac{13}{54}\tau_6-\frac{7}{54}\tau_3^2-\frac{247}{3456}\tau_2^3+\frac{793}{864}\tau_2\tau_4+\frac{7}{27}\tau_6'\,,\nonumber\\
    &\mathcal{P}(M_{10}''')=\frac{83408}{7}\tau_7+\frac{5553977}{9}\tau_3\tau_4-349336\tau_2\tau_5-\frac{376493}{36}\tau_3\tau_2^2\,.
\end{align}

\providecommand{\href}[2]{#2}\begingroup\raggedright\endgroup

\end{document}